\documentclass[useAMS,usenatbib,usegraphix]{mn2e}
\pdfoutput=1
\usepackage{times}
\usepackage{subfig}
\usepackage{aas_macros}
\usepackage{epsfig}
\usepackage{amssymb}
\usepackage{amsmath} 
\usepackage{hyperref}
\usepackage{fixltx2e}
\usepackage{tikz}
\usetikzlibrary{arrows,shapes,backgrounds}
\title[Modelling VV CrA]
{Understanding discs in binary YSOs -- detailed modelling of VV~CrA\thanks{Based on observations made with European Southern Observatory (ESO) telescopes at the La Silla Paranal Observatory under programs 74.C-0209(A), 75.C-0014(A) \& 91.C-0768(A)}}
\author[P. Scicluna et al.]
{P. Scicluna$^{1,2}$, S. Wolf$^1$, T. Ratzka$^3$, G. Costigan$^4$, R., Launhardt$^5$, C. Leinert$^5$, F. Ober$^1$, \and C.F. Manara$^{6}$,  L. Testi$^{7,8,9}$ \\
$^1$ITAP, Universit\"at zu Kiel, Leibnizstr. 15, 24118 Kiel, Germany\\
$^2$Academia Sinica, Institute of Astronomy and Astrophysics, Taipei 10617, Taiwan\\
$^3$Institute for Physics/IGAM, NAWI Graz, Karl-Franzens-Universit\"at, Universit\"atsplatz 5/II, 8010, Graz, Austria\\
$^4$Leiden Observatory, University of Leiden, PB 9513, 2300 RA, Leiden, The Netherlands\\
$^5$Max-Planck-Institut für Astronomie, K\"onigstuhl 17, 69117, Heidelberg, Germany \\
$^6$Scientific Support Office, Directorate of Science and Robotic Exploration, European Space Research and Technology Centre (ESA/ESTEC), Keplerlaan 1,\\ 2201 AZ, Noordwijk, The Netherlands \\
$^7$European Southern Observatory, Karl-Schwarzschild-Str. 2, D-85748 Garching b. M\"unchen, Germany\\
$^8$Excellence Cluster Universe, Boltzmannstr. 2, D-85748 Garching, Germany \\
$^9$INAF-Osservatorio Astrofisico di Arcetri, Largo E. Fermi 5, I-50125 Firenze, Italy\\
}
\date{Received:}

\newcommand{\Lsun}{L$_{\odot}$}
\newcommand{\Msun}{M$_{\odot}$}
\newcommand{\Rsun}{R$_{\odot}$}

\begin{document}
\maketitle

\begin{abstract}
Given that a majority of stars form in multiple systems, in order to fully understand the star- and planet-formation processes we must seek to understand them in multiple stellar systems.
With this in mind, we present an analysis of the enigmatic binary T-Tauri system VV Corona Australis, in which both components host discs, but only one is visible at optical wavelengths.
We seek to understand the peculiarities of this system by searching for a model for the binary which explains all the available continuum observations of the system.
We present new mid-infrared interferometry and near-infrared spectroscopy along with archival millimetre-wave observations, which resolve the binary at 1.3\,mm for the first time.
We compute a grid of pre-main-sequence radiative transfer models and calculate their posterior probabilities given the observed spectral energy distributions and mid-infrared interferometric visibilities of the binary components, beginning with the assumption that the only differences between the two components are their inclination and position angles.
Our best-fitting solution corresponds to a relatively low luminosity T-tauri binary, with each component's disc having a large scale height and viewed at moderate inclination ($\sim 50^\circ$), with the infrared companion inclined by $\sim 5^\circ$ degrees more than the primary.
Comparing the results of our model to evolutionary models suggests stellar masses $\sim 1.7$\,\Msun\,and an age for the system of 3.5\,Myr, towards the upper end of previous estimates.
Combining these results with accretion indicators from near-IR spectroscopy, we determine an accretion rate of 4.0 $\times 10^{-8}$\,\Msun\,yr$^{-1}$ for the primary.
We suggest that future observations of VV~CrA and similar systems should prioritise high angular resolution sub-mm and near-IR imaging of the discs and high resolution optical/NIR spectroscopy of the central stars.
\end{abstract}
 
\begin{keywords}
stars: pre-main-sequence -- binaries: general -- stars: variables: T Tauri -- protoplanetary discs -- accretion, accretion discs
\end{keywords}

\section{Introduction}\label{sec:int}

	Stars form in turbulent molecular clouds, and as a result of turbulence, regions of the cloud condense to form cores, which may then continue to collapse under the influence of gravity \citep[e.g.][]{2014prpl.conf...27A}.
	This may result in the formation of a single star, or the core may fragment and form many stars \citep{1984Ap&SS..99...41Z}.
	Angular momentum must be conserved during this collapse, inevitably resulting in any stars being surrounded by a disc of material, which may then be accreted by the star, ejected from the system, or potentially condense into a planetary system \citep{2011ARA&A..49...67W}.
	In the event that two or more stars form on gravitationally bound orbits, they will form a binary or multiple protostellar system \citep{1984Ap&SS..99...41Z}.
	
	Protostellar and pre-main-sequence binaries provide unique insights into the star-formation process \citep{1984Ap&SS..99...41Z,2009ApJ...704..531K,duchenekraus2013}.
	Although more complex, it is known that many young stars form as binary or multiple systems. 
	The probability that a companion is present depends on the mass of the primary and the environment in which the stars form \citep[for details see][and references therein]{duchenekraus2013}. 
	Thanks to their quasi-coeval nature, binary systems provide key tests for models of stellar evolution; pre-main sequence binaries are therefore crucial to the calibration of evolutionary models at these early stages \citep{2009ApJ...704..531K}.
	Of particular interest are systems where one or both components still host protoplanetary discs, enabling tests of the evolution of the discs and the star-disc interaction, and in which the binary separation is large, so that the evolution is not complicated by binary interactions \citep{2011ARA&A..49...67W}.
	Furthermore, studying coeval systems could yield new constraints for models of planet formation.
	For example, stars in multiple systems are less likely to be observed to have a disc at any age when compared with single stars \citep{2009ApJ...696L..84C,2013A&A...554A..43D,2015arXiv151105965D}, yet planets have been detected in a variety of multiple systems \citep[e.g.][]{2011Sci...333.1602D,2012A&A...542A..92R}. 
	
	VV Corona Australis (VV~CrA) is a young binary system in the Corona Australis star-forming region. 
	Situated approximately 20 arc minutes south east of the prominent Coronet cluster, it is the brightest component in what appears to be a smaller condensation of young stars \citep{2013A&A...551A..34S}.
	Previously thought to be a single star, according to \citet{1993A&A...278...81R}, a companion was discovered approximately 2$^{\prime\prime}$ to the north east of the optical source by Frogel (unpublished) due to a displacement of the near-infrared peak from the optical coordinates. 
	At the time, this companion dominated the near-infrared emission of the system despite being too extinguished to be visible at shorter wavelengths \citep{1995A&AS..114..135C}.
	This infrared companion, now referred to as VV~CrA~NE, has since faded \citep{przygodda2004} such that the optical component (VV CrA SW) is presently the brighter source at all wavelengths at which the binary can be resolved \citep{2011ApJ...729..145K}.
	Although the mass ratio of the binary is unknown, throughout this paper we follow \citet{2009ApJ...701..163S} in referring to the optical component VV~CrA~SW as the primary.

	The peculiarities of the binary have made it an intriguing observational target, however it remains enigmatic.
	The combination of high extinction and ongoing accretion makes the determination of the spectral types of the stars nearly impossible, since veiling fills in many of the photospheric lines \citep{2014ApJ...786...97H}.
	Nevertheless, many values have been published for the primary, ranging from K7 \citep{1986A&AS...64...65A} to K1 \citep{2000A&AS..146..323N}.
	
	Furthermore, the origins of both the high obscuration and the variability of the secondary remain unknown, although two viable solutions have been proposed by \citet{2009ApJ...701..163S}.
	In one solution (their case B) VV~CrA~NE is inclined such that we view it through its own circumstellar disc, while the second (case A) requires that the disc of the primary is the source of the obscuration.
	In either case, the variability may be caused either by column density variations leading to changes in the extinction, or by an accretion outburst similar to that of FU Orionis \citep{2009ApJ...701..163S}.
	
	However, some progress has been made thanks to observations across a wide wavelength range.
	Both sources are variable in the infrared, although the variability of the primary is much weaker than that of the secondary.
	Both show emission in a variety of spectral lines \citep{2003ApJ...584..853P} and appear to be driving outflows \citep{2003A&A...397..675T}.
	Both components are bright in the (sub-)mm, and \citet{2010A&A...515A..77L} used unresolved SMA data to determine a total dust mass of $3.9 \times 10^{-4}$\,\Msun for the binary, and a spectral index indicative of significant grain growth.

	In this paper, we present a selection of new observations along with a re-reduction of archival data, which are then used, along with data from the literature, to constrain radiative transfer modelling.
	In \S\ref{sec:obs} we present new infrared interferometric and spectroscopic data along with archival sub-millimetre observations.
	\S\ref{sec:mod} gives an overview of the methods used for the radiative transfer calculations and to infer the best-fitting model, the results of which are discussed in \S\ref{sec:dis}.
	We then discuss observational priorities that will enable future modelling efforts to improve upon our results, before summarising our conclusions in \S\ref{sec:conc}.

\section{Observations}\label{sec:obs}
	\subsection{MIDI visibilities}\label{sec:midi}

	\begin{table*}
\caption{Journal of MIDI Observations. The length and position angle of the projected baseline has been determined from the fringe tracking sequence, the airmass from the photometric frames.}
\label{journal}
\centering
\begin{tabular}{ccccccccccc}
\hline\hline
Date of     & Universal     & Object	 & IRAS & \multicolumn{2}{c}{proj.~Baseline} &  AM & Interferometric & Photometric & Flags\\
Observation & Time          &		 & [Jy] & \     [m] & [deg]                  & 	   & Frames	     & Frames	   & \\
\hline
\noalign{\smallskip}
02-06-2004  & 00:33 - 00:55 & {HD 112213} &   10.8 & 45.4 & 40.8 & 1.1 & \ 8000$\times$12\,ms & 2$\times$1500$\times$12\,ms & $^s$ \\
02-06-2004  & 06:59 - 07:27 & {VV~CrA~SW} &        & 45.7 & 42.7 & 1.0 & \ 8000$\times$15\,ms & 2$\times$1500$\times$15\,ms &      \\
02-06-2004  & 07:33 - 07:48 & {HD 178345} &\ \ 8.6 & 44.9 & 45.1 & 1.0 & \ 8000$\times$15\,ms & 2$\times$1500$\times$15\,ms & $^s$ \\
02-06-2004  & 07:57 - 08:40 & {VV~CrA~SW} &        & 43.0 & 51.0 & 1.1 & \ 8000$\times$15\,ms & 2$\times$1500$\times$15\,ms & \\
02-06-2004  & 10:14 - 10:31 & {HD 178345} &\ \ 8.6 & 35.2 & 61.3 & 1.4 & \ 8000$\times$15\,ms & 2$\times$1500$\times$15\,ms & $^s$ \\
\noalign{\smallskip}
\hline
\noalign{\smallskip}
04-11-2004  & 00:01 - 00:26 & {HD 178345} &\ \ 8.6 & 57.0 & 145.7 & 1.4 &  20000$\times$12\,ms & 2$\times$5000$\times$12\,ms & $^s$ \\
04-11-2004  & 01:07 - 01:31 & {VV~CrA~SW} &        & 52.7 & 164.8 & 2.0 &  12000$\times$12\,ms & 2$\times$3000$\times$12\,ms &	\\
04-11-2004  & 01:31 - 02:10 & {VV~CrA~NE} &        & 51.9 & 175.9 & 2.6 &  12000$\times$12\,ms & 2$\times$3000$\times$12\,ms & $^{e,v}$	\\
04-11-2004  & 02:10 - 02:47 & {HD 188603} &   11.5 & 45.5 & 168.6 & 2.5 &  16000$\times$12\,ms & 2$\times$5000$\times$12\,ms & $^v$ \\
04-11-2004  & 02:47 - 03:54 & {HD 25604}  &\ \ 5.1 & 60.7 & 117.1 & 1.7 &  16000$\times$12\,ms & 2$\times$5000$\times$12\,ms &	\\
04-11-2004  & 05:04 - 05:29 & {HD 20644}  &   14.7 & 59.1 & 101.5 & 1.7 &  12000$\times$12\,ms & 2$\times$3000$\times$12\,ms &	\\
04-11-2004  & 07:18 - 07:36 & {HD 37160}  &\ \ 6.5 & 61.0 & 107.4 & 1.2 & \ 8000$\times$12\,ms & 2$\times$2000$\times$12\,ms & $^s$ \\
04-11-2004  & 09:00 - 09:23 & {HD 50778}  &   17.3 & 61.0 & 112.6 & 1.0 & \ 8000$\times$12\,ms & 2$\times$2000$\times$12\,ms &	\\
\noalign{\smallskip}
\hline
\noalign{\smallskip}
29-05-2005  & 06:19 - 07:00 & {HD 139127} &   12.3 & 82.8 &  51.6 & 1.3 & \ 8000$\times$18\,ms & 2$\times$4000$\times$18\,ms & $^{p,s}$ \\
29-05-2005  & 07:28 - 07:56 & {HD 142198} &\ \ 6.3 & 92.0 &  41.8 & 1.6 & \ 8000$\times$18\,ms & 2$\times$4000$\times$18\,ms &	   \\
29-05-2005  & 08:54 - 09:17 & {HD 152820} &\ \ 7.6 & 77.7 &  49.7 & 1.7 & \ 8000$\times$18\,ms & 2$\times$4000$\times$18\,ms &	   \\
29-05-2005  & 09:17 - 09:56 & {VV~CrA~NE} &        & 88.8 &  47.5 & 1.2 & \ 8000$\times$18\,ms & 2$\times$4000$\times$18\,ms &	   \\
29-05-2005  & 09:56 - 10:19 & {VV~CrA~SW} &        & 85.7 &  49.1 & 1.3 & \ 8000$\times$18\,ms & 2$\times$4000$\times$18\,ms &	   \\
29-05-2005  & 10:19 - 10:53 & {HD 178345} &\ \ 8.6 & 81.6 &  51.4 & 1.4 & \ 8000$\times$18\,ms & 2$\times$4000$\times$18\,ms & $^{p,s}$ \\
\noalign{\smallskip}
\hline
\noalign{\smallskip}
30-05-2005  & 22:35 - 23:15 & {HD 102839} &\ \ 5.9 & 58.9 &  91.6 & 1.4 & \ 8000$\times$18\,ms & 2$\times$4000$\times$18\,ms &	     \\
30-05-2005  & 00:17 - 00:41 & {HD 102461} &\ \ 8.6 & 62.0 & 114.0 & 1.2 & \ 8000$\times$18\,ms & 2$\times$4000$\times$18\,ms & $^{s,v}$   \\
30-05-2005  & 01:24 - 01:56 & {HD 139127} &   12.3 & 54.1 &  88.7 & 1.2 & \ 8000$\times$18\,ms & 2$\times$4000$\times$18\,ms & $^{p,s}$   \\
30-05-2005  & 01:56 - 02:09 & {HD 139127} &   12.3 & 56.1 &  92.1 & 1.1 & \ 8000$\times$18\,ms & 2$\times$4000$\times$18\,ms & $^{a,s,v}$ \\
30-05-2005  & 03:18 - 03:45 & {HD 178345} &\ \ 8.6 & 41.6 &  73.1 & 1.5 & \ 8000$\times$18\,ms & 2$\times$4000$\times$18\,ms & $^s$       \\
30-05-2005  & 03:45 - 03:58 & {HD 178345} &\ \ 8.6 & 43.5 &  75.9 & 1.4 & \ 8000$\times$18\,ms & 2$\times$4000$\times$18\,ms & $^{a,s}$   \\
30-05-2005  & 03:58 - 04:29 & {VV~CrA~NE} &        & 48.7 &  84.8 & 1.3 & \ 8000$\times$18\,ms & 2$\times$4000$\times$18\,ms &	     \\
30-05-2005  & 04:29 - 04:42 & {VV~CrA~NE} &        & 50.4 &  87.0 & 1.2 & \ 8000$\times$18\,ms & 2$\times$4000$\times$18\,ms & $^{a,e}$       \\
30-05-2005  & 04:42 - 05:04 & {VV~CrA~SW} &        & 53.0 &  90.3 & 1.2 & \ 8000$\times$18\,ms & 2$\times$4000$\times$18\,ms &	     \\
30-05-2005  & 05:04 - 05:16 & {VV~CrA~SW} &        & 54.4 &  92.1 & 1.1 & \ 8000$\times$18\,ms & 2$\times$4000$\times$18\,ms & $^a$       \\
30-05-2005  & 05:16 - 05:28 & {VV~CrA~SW} &        & 55.7 &  93.9 & 1.1 & \ 8000$\times$18\,ms & 2$\times$4000$\times$18\,ms & $^a$       \\
30-05-2005  & 05:28 - 05:49 & {VV~CrA~NE} &        & 57.6 &  96.9 & 1.1 & \ 8000$\times$18\,ms & 2$\times$4000$\times$18\,ms &	     \\
30-05-2005  & 05:49 - 06:14 & {HD 133774} &\ \ 8.4 & 49.4 & 130.6 & 1.3 & \ 8000$\times$18\,ms & 2$\times$4000$\times$18\,ms & $^s$       \\
30-05-2005  & 07:01 - 07:35 & {HD 139127} &   12.3 & 57.6 & 149.7 & 1.5 & \ 8000$\times$18\,ms & 2$\times$4000$\times$18\,ms & $^{p,s}$   \\
30-05-2005  & 08:08 - 08:31 & {HD 152820} &\ \ 7.6 & 54.5 & 144.1 & 1.4 & \ 8000$\times$18\,ms & 2$\times$4000$\times$18\,ms &	     \\
30-05-2005  & 08:55 - 09:21 & {HD 164064} &\ \ 5.0 & 42.2 & 124.9 & 1.5 & \ 8000$\times$18\,ms & 2$\times$4000$\times$18\,ms &	     \\
30-05-2005  & 09:21 - 10:26 & {VV~CrA~NE} &        & 56.7 & 142.8 & 1.3 & \ 8000$\times$18\,ms & 2$\times$4000$\times$18\,ms &	     \\
30-05-2005  & 10:26 - 10:59 & {HD 178345} &\ \ 8.6 & 56.7 & 147.2 & 1.5 & \ 8000$\times$18\,ms & 2$\times$4000$\times$18\,ms & $^{p,s,v}$ \\
\noalign{\smallskip}
\hline
\noalign{\smallskip}
\multicolumn{8}{l}{$^{a}$ no recentering of the beams performed}\\
\multicolumn{8}{l}{$^{e}$ excluded from fitting}\\
\multicolumn{8}{l}{$^{p}$ photometry of one or both beams repeated}\\
\multicolumn{8}{l}{$^{s}$ spectrophotometric calibrator}\\
\multicolumn{8}{l}{$^{v}$ peculiar instrumental visibility}\\

\noalign{\smallskip}
\hline

\end{tabular}
\end{table*}

	VV CrA was observed in June 2004, November 2004, and May 2005 (Tab.~\ref{journal}) with the Mid-infrared Interferometric (MIDI) instrument \citep{2003Ap&SS.286...73L,2003SPIE.4838..893L,2004SPIE.5491.1666M} at the VLTI, as part of programmes ESO-074.C-0209(A) and ESO-075.C-0014(A) (P.I. Leinert). 
	Both components of the binary were individually measured due to their comparatively large separation. 
	The baselines UT1-UT3, UT2-UT3, and UT3-UT4 were used and the collected light was dispersed with the prism ($R\sim 30$).

    The data were reduced with the MIA+EWS package\footnote{\url{http://www.strw.leidenuniv.nl/~nevec/MIDI/}}. 
    MIA is based on the analysis of the power spectrum, while EWS implements a coherent analysis of the interferometric fringe signal (Jaffe et al., 2004). 
    The visibilities used for the modelling have been derived with MIA and confirmed with EWS. 
    All visibilities have been calibrated by all the calibrators taken in the same night. 
    The stability of the transfer function is reflected by the errors shown together with the visibilities. 
    Calibrators with peculiar visibilities have been ignored and are indicated in the journal of observations (Tab.~\ref{journal}).
    
    For technical reasons, the single dish spectra derived from MIDI measurements vary over time in absolute calibration.
    Therefore, they are scaled to the better calibrated TIMMI2 spectra from \citet{przygodda2004}

    We also include three older measurements already obtained with MIDI in June 2003 \citep{przygodda2004}. 
    These observations have been made with projected baseline lengths between 99.7\,m and 102.2\,m along position angles between 35.7$^{\circ}$ and 23.9$^{\circ}$. 

\begin{figure*}
		\subfloat[]{\includegraphics[scale=0.5,clip=true,trim=1.cm 1.cm 1.5cm 10cm]{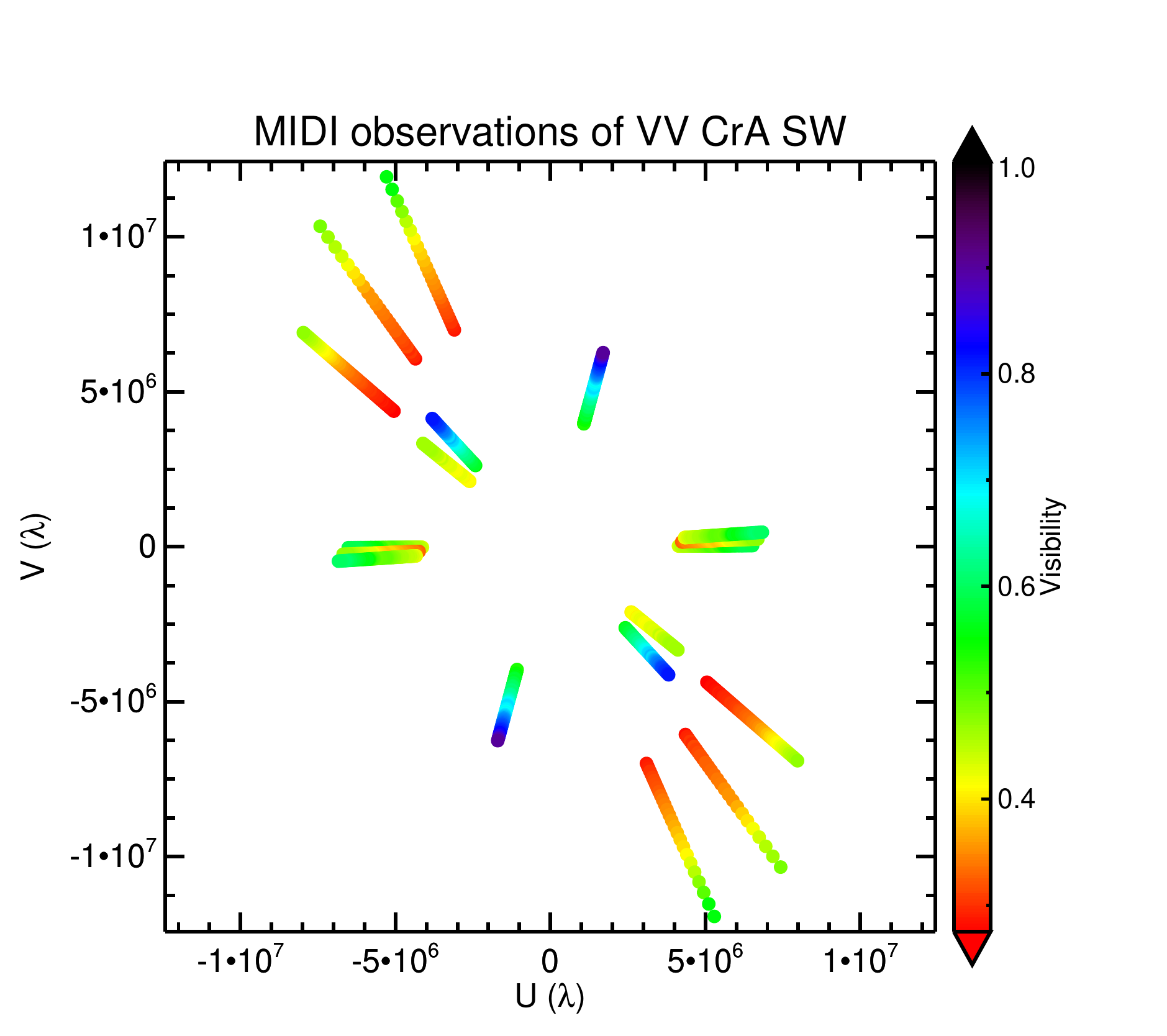}}	
	    \subfloat[]{\includegraphics[scale=0.5,clip=true,trim=1.cm 1.cm 1.5cm 10cm]{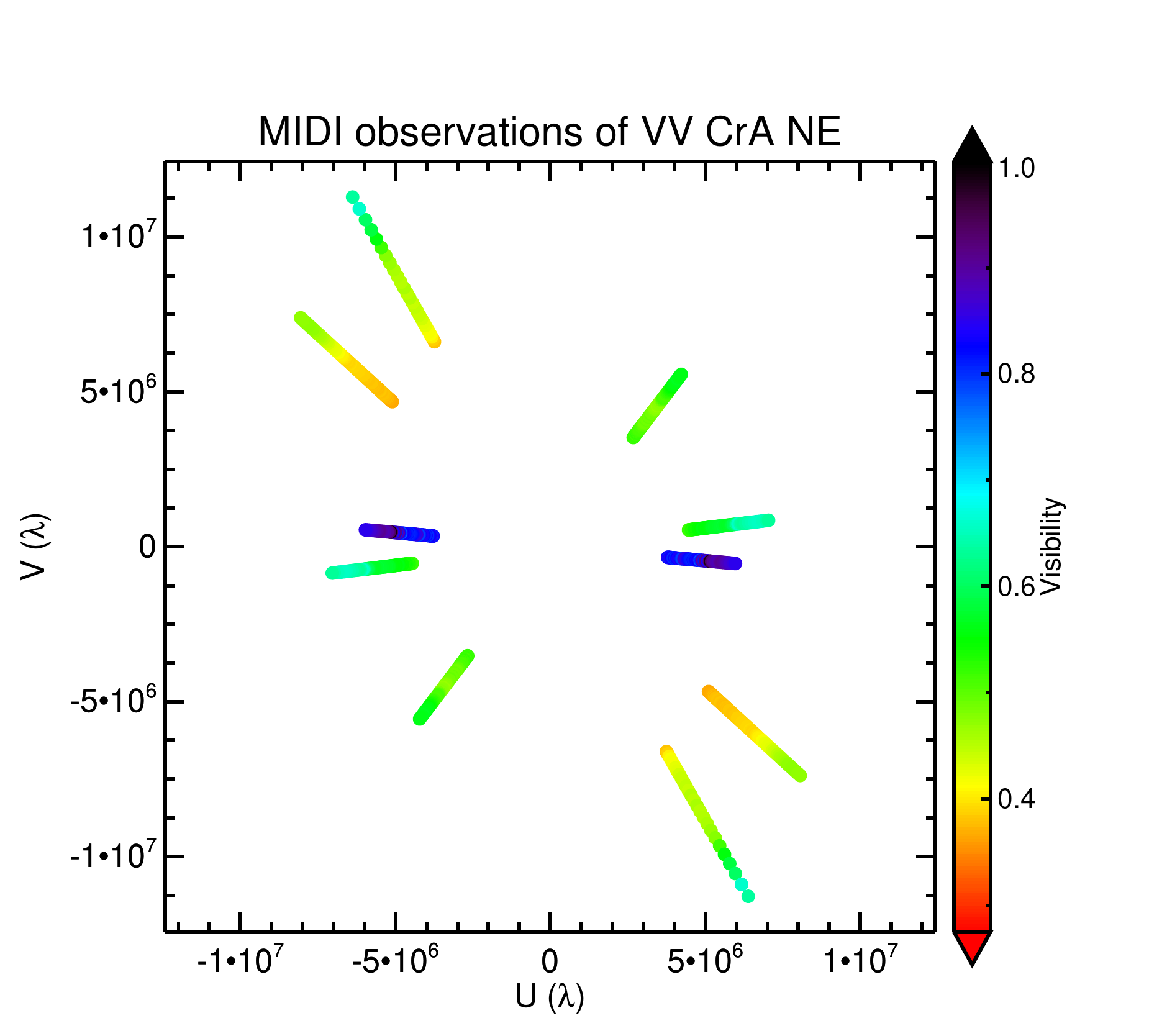}}	
	    \caption{$a)$ Wavelength-resolved visibilities plotted in the UV plane for the observations of VV~CrA~SW. $b)$ As $a)$ but for the 5 observations of VV~CrA~NE. UV distances are shown in units of the wavelength; as a result, each line has the longest wavelengths toward the centre and shortest wavelengths toward the edge of the plot.}
	    \label{fig:midi}
	\end{figure*}

	The resulting visibilities are plotted in Fig. \ref{fig:midi}.
	Two observations of VV~CrA~NE are excluded from the plot and from the model fitting (\S\,\ref{sec:mod}); these are flagged in Tab.~\ref{journal} as $e$.
	
    \subsection{CRIRES Spectra}\label{sec:spec}
    VV~CrA~SW was observed three times over the nights 22,23,24-AUG-2013 as part of programme ESO-091.C-0768(A) (P.I. Costigan) with the CRIRES instrument at the ESO-VLT \citep{2004SPIE.5492.1218K}. 
    
    The observations were performed with a central reference wavelength of 2170.573\,nm and slit width of 0.4\," giving a spectral resolution $R = \frac{\lambda}{\Delta\lambda}\sim$ 50,000. In order to provide good sky subtraction, the star was nodded along the slit between individual exposures. Three sub-integrations of 30 seconds were taken for each observation. Standard reduction steps for infrared observations were performed using IRAF, including flat-fielding, bias-subtraction and spectral extraction. Due to the nodding on the slit, each observation block results in two spectra which were then median combined. These spectra were then telluric corrected with standards observed on the night. 
    
   \begin{figure}
	    \centering
    	\resizebox{\hsize}{!}{\includegraphics[width=\textwidth]{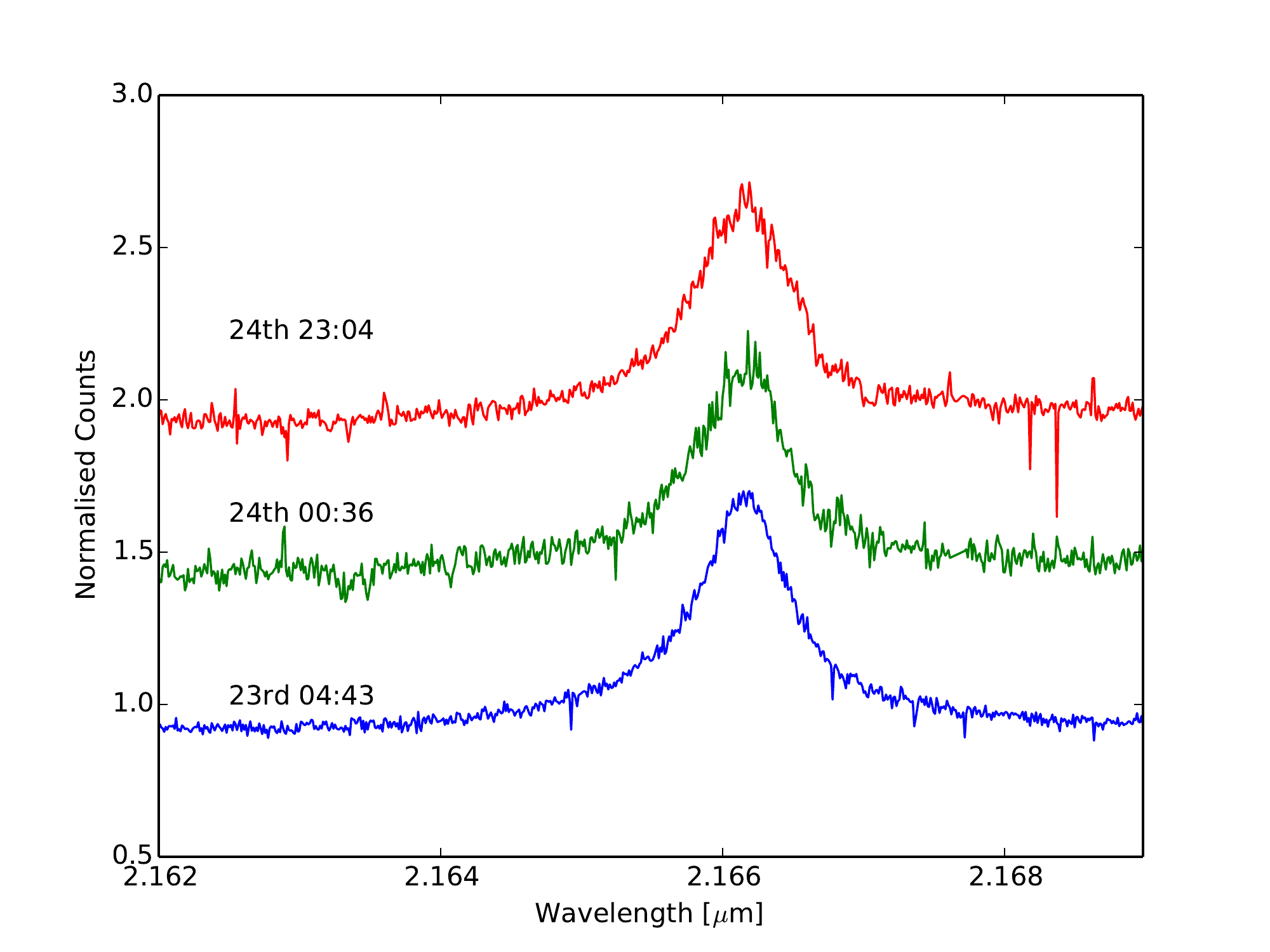}}
    	\caption{Median--combined CRIRES spectra from each observing night, showing the region of the spectra covering the Br-$\gamma$ line. }\label{fig:brg}
	\end{figure}  
    
    The wavelength range covered by these observations is $\sim$ 2143\,-\,2195\,nm. Unfortunately, there were no photospheric lines visible within these spectra, and so a spectral type derivation was not possible. However, the Br-$\gamma$ emission line lies within this range, and is closely associated with accretion \citep{1998AJ....116.2965M,2004AJ....128.1294C,2011A&A...534A..32A}.  No significant variations in the equivalent width of this line were measured over the course of the three nights, with a mean measurement of 9\,$\pm$~0.7\,\AA. 
 
	Using the observed NIRC K-band flux of 1.63\,Jy for this target (Koresko, priv. comm.), corrected for extinction using the redding law by \citet{1991Ap&SS.184....9S}, we converted the Br-$\gamma$ EW into a luminosity (taking the distance to VV~CrA~SW to be 130\,pc, \citealt{1981AJ.....86...62M}) which was found to be $7.2\times~10^{-4}$\,\Lsun. Using the \citet{2014A&A...561A...2A} accretion luminosity relation for  Br-$\gamma$, this then gives an accretion luminosity of 0.81~$\pm$~0.07~\Lsun. Here the spread comes from considering the spread in the measured EW over the course of the three days. 
	
	Following the methods of \citet{2014arXiv1409.2831S} we also tested for the presence of any extended emission from the Br-$\gamma$ line using spectro-astrometry. The original observations were not designed to probe the surrounding structure, and so only contained one position angle. The only conclusion we can so derive from these observations is that there is no signal of extended emission at the 1-10\,mas level in the Br-$\gamma$ line in the north-south direction. 
 
 \subsection{Millimetre continuum data}   \label{sec:mm}
    \subsubsection{SMA data}\label{sec:sma}
    	In order to improve the constraints on the emission at long wavelengths, we retrieved archival data taken using the Submillimetre Array at 1.3\,mm.
    	VV~CrA was observed on 01-Oct-2008 in Compact configuration with 6 antennae \citep[previously published in][]{2010A&A...515A..77L} and on 12-Sept-2012 \& 18-Sept-2012 in Extended configuration  with 7 antennae (unpublished).
    	Both datasets were reduced separately and in combination to explore the loss of flux on large scales.
    	
\begin{figure}
    	\resizebox{\hsize}{!}{\includegraphics[scale=0.5,clip=true,trim=1.cm 14.cm 3.5cm 2cm]{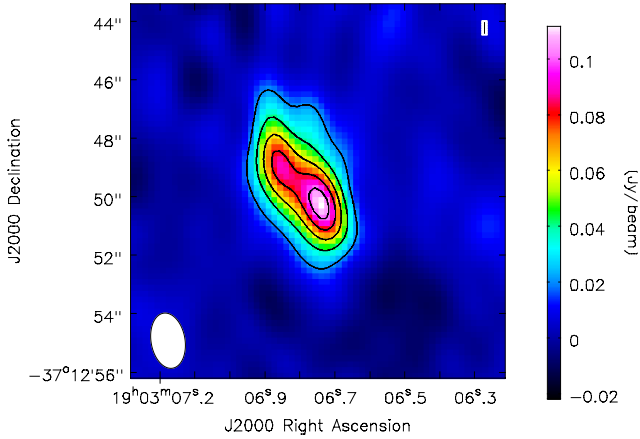}}
		\caption{Final reduced continuum image reconstructed from the SMA observations at 1.3\,mm. The white oval shows the size and shape of the synthesised beam.} 
		\label{fig:sma}
    	\end{figure}

    	The calibrated data was downloaded from the SMA archive and imaged using {\sc CASA}.
    	After masking edge channels and checking for the presence of spectral lines, the data were imaged interactively using the \textit{clean} routine with natural weighting. 
    	The final reduced image, using both array configurations, is shown in Fig \ref{fig:sma}. 
    	Clearly resolved peaks are visible at approximately the same positions as the infrared locations of the two sources. 
    	In order to extract continuum fluxes, 2D gaussians were fitted simultaneously to both components using the \textit{imfit} routine with the literature positions of the two components as initial guesses. 
    	Images reconstructed using only the extended array observations resolve both components of the binary more clearly, separating the two peaks more distinctly, however more than 50\% of the single-dish flux detected by \citet{2003A&A...409..235C} is missing. 
    	We therefore do not show the image or use it to derive fluxes.
    	Folding in the observations taken in the compact configuration reduces this missing-flux problem somewhat, although there is still $\sim 25\%$ missing flux.
    	This suggests that either circumbinary structure or structure associated with the parent cloud are being filtered out; future observations using ALMA including the compact array are required to identify the source of this emission.
    	
\subsubsection{ATCA observations}    	
Prior to the 2008 observations presented by \citet{2010A&A...515A..77L}, VV~CrA was observed with ATCA at 3\,mm on 2005-08-14 in configuration H75 and on 2005-08-17 in H214, with good weather conditions for both observations.
Combining the two configurations provided physical baselines ranging from 
22 to 247~m. All antennae were equipped with cooled SIS receivers, which 
provided average system temperatures of 200--350~K at the observing frequency. 
A digital correlator was used with two 128 MHz wide spectral windows centered at 
93.504 and 95.552~GHz, respectively. Combining the two sidebands resulted in an 
effective observing wavelength of 3.17~mm. The primary beam size at this wavelength was $\sim$\,38$^{\prime\prime}$.
Amplitude and phase were calibrated through frequent observations (typically 
every 20 minutes) of the nearby quasar 1921-293. The absolute flux density was
calibrated using the secondary calibrator 1253-055 in each track, the flux of which
was regularly compared to Uranus and adopted as 14.7\,Jy for the time of the 
observations. Additional effort was made to improve the gain-elevation calibration 
of the antennae, which can significantly affect the flux density scale, especially when observing at high elevation. The resulting total flux uncertainty is estimated 
to be $\sim$~20\%.

The data were calibrated and images produced from the combined uv data using MIRIAD (Sault et al. 1995) and its CLEAN algorithm. 
Of the different uv-weighting schemes explored, the image produced with \emph{robust} uv-weighting parameter -0.5 (Briggs et al. 1999) was finally selected as the optimum. 
The resulting synthesized FWHM beam size is  2.39$^{\prime\prime}$\,$\times$\,1.73$^{\prime\prime}$ (PA 86.7$^{\circ}$). 
The 1 sigma noise level in the final map is $\sim$~1~mJy/beam.

\begin{figure} 
\resizebox{\hsize}{!}{
\begin{tikzpicture}
\node[anchor= south west, inner sep=0] (image) at (0,0) {\includegraphics[clip=true,trim=1.cm 16.cm 3.5cm 2cm]{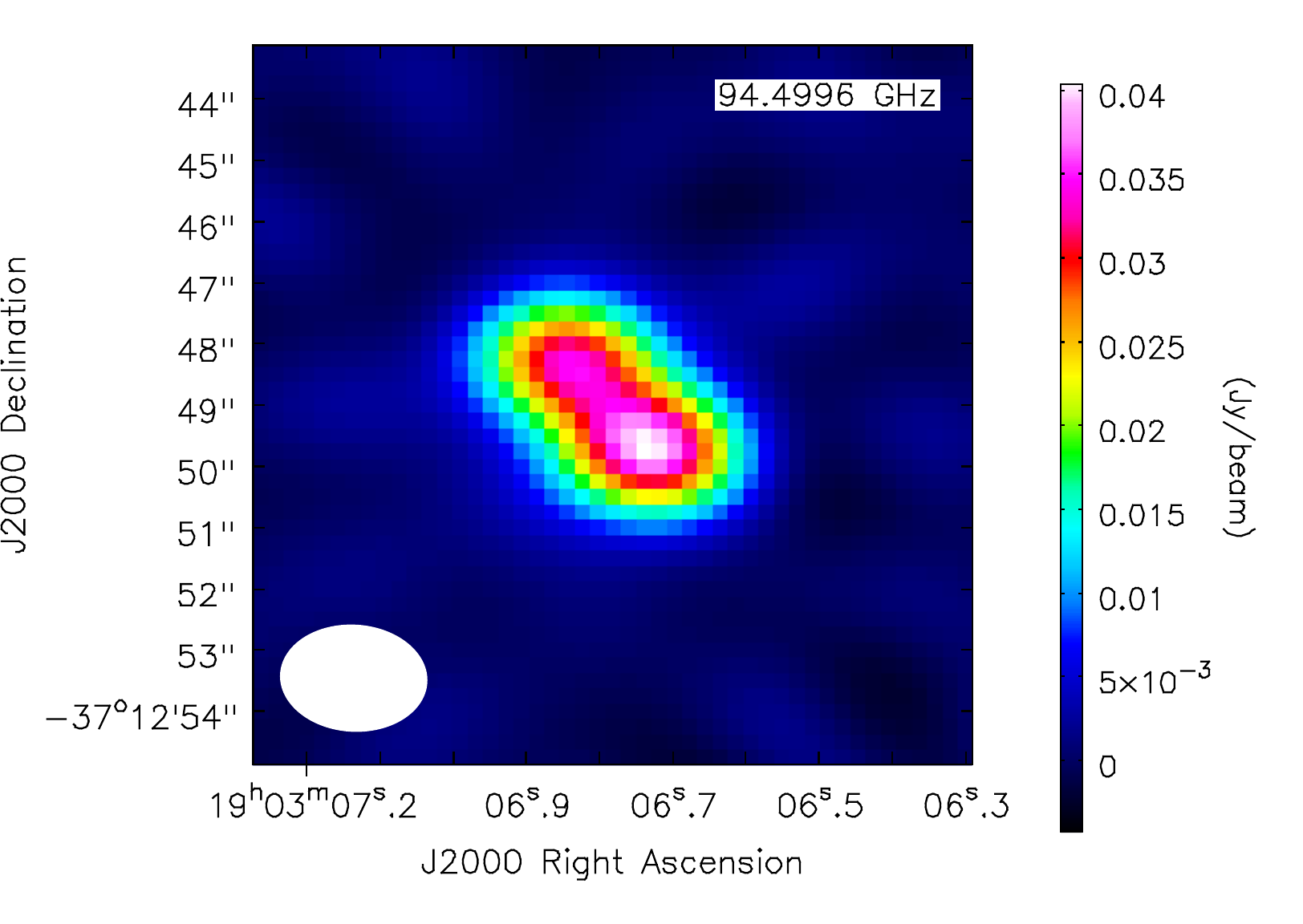}};
\begin{scope}[x={(image.south east)},y={(image.north west)}]
	\node[star,star point ratio=3,minimum width=1mm,scale=0.4,draw] at (0.5,0.52) {S}; 
	\node[star,star point ratio=3,minimum width=1mm,scale=0.4,draw] at (0.44,0.6) {S};	 
\end{scope}
\end{tikzpicture}
}	
\caption{Reconstructed image from the 2005 ATCA dataset. The approxiate locations of the two components are marked by stars. }\label{fig:3mm}
\end{figure}

The resulting image (Fig.~\ref{fig:3mm}) clearly resolves the two sources, although the individual 
sources (disks) remain unresolved and separating disk emission from possible 
envelope or circumbinary contributions is not feasible. 
Image plane fitting of 2-D gaussians results in integrated fluxes of 40~$\pm$~8 mJy
and 35~$\pm$~8 mJy for the SW and NE components, respectively. The total flux 
density of 75~$\pm$~15 mJy was consistently measured from both the uv data and 
the image plane. This value is consistent with the total flux of 69.5~$\pm$~14 mJy 
(at the same frequency) reported by \citet{2010A&A...515A..77L} from their 
measurements on 2008-08-02, although the fluxes they attribute to the individual 
components are about 25\% lower than ours (albeit with similar flux ratio). These 
slight discrepancies between individual contributions and total flux could both be 
a result of their missing short baselines (H214 config only) or our uncertainties 
in separating disk from extended emission. However, \citet{2010A&A...515A..77L} report 
for 2008-08-03, i.e., only one day later, a total flux density of only 44.2 +/- 9 mJy.
Although we cannot exclude that this points to a real short-term flux variability, the consistency between their 
2008-08-02 flux and our 2005 flux makes this unlikely.

The 3\,mm fluxes are not included in our subsequent model fitting for a number of reasons.
First and foremost, it is not possible to uniquely attribute the differences between our fluxes and the \citet{2010A&A...515A..77L} fluxes to either measurement uncertainty or real variability.
Secondly, as discussed below, we cannot exclude a significant contribution from free-free emission at 3\,mm, while the model only treats dust emission.

    	Using these new resolved 1.3\,mm and 3\,mm fluxes
    	we derive spectral indices $\alpha_{\rm mm}$ of 1.9\,$\pm$\,0.4 and 1.8\,$\pm$\,0.4 for the primary and secondary, respectively. 
    	These can be compared with values of $\alpha_{\rm mm}$ of 2.2\,$\pm$\,0.5 and 2.3\,$\pm$\,0.6 when using our 1.3\,mm fluxes and the \citet{2010A&A...515A..77L} 3\,mm fluxes; both values are mutually consistent as expected.
    	This indicates either substantial grain-growth, with a sizeable population of grains with radius $>$\,1\,mm, that the discs are very optically thick at mm-wavelengths, or a significant contribution from free-free emission at 3\,mm.
    	However, further observations are required to attribute this to a particular mechanism.
    
\section{Modelling}\label{sec:mod}
	\subsection{Radiative transfer modelling with {\sc MC3D}}\label{sec:mc3d}
		{\sc MC3D} \citep{1999A&A...349..839W,2003CoPhC.150...99W} solves the radiative transfer equation self-consistently using Monte Carlo methods.
		Optimised for dusty circumstellar discs, it randomly propagates packets of radiative energy (photons) through the medium. 
		Each packet is monochromatic and of fixed energy.
		Packets interact randomly with the medium, either through scattering events - in which only the direction and stokes vector of the photon change - or absorption \& re-emission events - where the wavelength of the re-emitted photon is likely to change.
		For a given dust-density distribution, {\sc MC3D} calculates the temperature distribution, and scattered and re-emitted fluxes.
		
		The temperature distribution is calculated assuming that the dust is in local thermodynamical equilibrium (LTE) and that the dust is heated only by the central star. 
		Once the dust-temperature distribution has been calculated, the spectrum of thermal re-emission can be calculated.
		For computational efficiency, this is done using ray-tracing techniques.
		Rays are generated ensuring that they sample all cells in the grid, stepped through the model grid to integrate the flux along each line of sight, and projected onto a detector.
		As this method only considers the contribution of thermal emission by dust, we also compute the contribution of the stellar radiation and scattered flux through monochromatic radiative transfer, in which the star launches radiation packets at each wavelength of interest, which are followed until they exit the model space.
		Images are computed using the same methods, with the results resolved into a number of pixels.
		
		\subsubsection{Disc density distribution}
			
			Although the radiative transfer is performed in 3 dimensions, to reduce the computational time we consider only 2-dimensional, axi-symmetric disc density distributions, which take the form
			\begin{equation}
			\rho\left(R,z\right) = \rho_0 \left(\frac{R}{R_0}\right)^{-\alpha} exp\left(-\frac{1}{2}\left(\frac{z}{h\left(R\right)}\right)^2\right), \label{eq:rhodisc}
			\end{equation}
			where the scale-height $h$ is given by
			\begin{equation}
			h\left(R\right) = h_0\left(\frac{R}{R_0}\right)^\beta ,
			\end{equation}
			where $R_0=$\,100AU, based on the $\alpha$-viscosity disc of \citet{1973A&A....24..337S}.
			The above density distribution is defined in the range $R_{\rm in}\leq R \leq R_{\rm out}$, the inner and outer radii of the disc.
			For a vertically isothermal gaseous disc in hydrostatic equilibirum $\alpha=3\left(\beta - 0.5\right)$ \citep{1973A&A....24..337S}; given that we only treat small grains (see \S\,\ref{sec:dust}) which should be dynamically coupled to the gas, we opt to define our density distribution only in terms of $\beta$. 
			This implicitly assumes that the disc is in hydrostatic equilibrium, although we \emph{do not} compute the equilibrium structure of the disc. 
			This distribution is then normalised to yield the desired total dust mass M$_{\rm d}$.
			The disc is heated by a star whose spectrum is that of a blackbody $B\left(\mathrm{T}_\ast\right)$ scaled to contain the required total stellar luminosity L$_\ast$.
			The temperature only has a significant effect on the emergent flux at optical wavelengths, because the IR emission is determined by the energy balance of the dust, which is dominated by the input luminosity.
			Similarly, we do not include any accretion luminosity or UV excess, as the optical and UV coverage of the SEDs are too sparse to be able to constrain this.
			This approach has been successfully applied in a number of studies of protoplanetary discs, for example \citet{2009A&A...505.1167S,2012A&A...543A..81M,2013A&A...553A..69G}.
		
		\subsubsection{Dust model}\label{sec:dust}

			Although there are signs of grain-growth in millimetre-wave observations of VV~CrA (\citealt{2010A&A...515A..77L} and \S\,\ref{sec:mm}) only ISM-like dust grains are included in the models.
			The energy balance of the disc is primarily determined by the more numerous small grains which absorb the stellar radiation more efficiently \citep{2014prpl.conf..339T}.
			These grains therefore dominate the short--wavelength emission (and hence the MIDI observations) and the total energy absorbed by dust; this in turn determines the integrated energy of the SED.
			However, large ($\sim 1$\,mm) grains emit more efficiently at mm\,wavelengths, and therefore dominate the appearance of (sub--)mm images and the spectral index.
			
			Theoretically, one expects that these larger dust grains should decouple from the gas density distribution, settling toward the mid-plane of the disc, while gas-drag would cause them to drift toward the inner regions of the disc \citep{2004A&A...421.1075D}.
			In the absence of multi-wavelength, spatially-resolved observations at millimetre wavelengths it is impossible to quantify the distribution of these larger dust grains.
			Since the MIDI observations are sensitive only to the surface layers of the inner regions of the disc, which are expected to be populated by smaller grains, only grains with sizes between 5\,nm and 250\,nm are included with a power-law size distribution ($q=-3.5$) as in \citet{1977ApJ...217..425M}, consisting of 62.5\% amorphous silicates and 37.5\% graphite, using optical constants from \citet{2001ApJ...548..296W}, and are assumed to be compact spheres.
			The cross-sections are computed using Mie calculus and averaged over the size distribution and dust compositions.
	
	\subsection{The fitting process}\label{sec:fit}
	
	\begin{table}
	\caption{Parameter space covered by RT models}
	\label{tab:gpars}
	\centering
	\begin{tabular}{l l l l l}
	\hline\hline
	Parameter & Range & Step & & Prior\\ \hline
	 & \\
	D (pc) & 130 & fixed  & &  \\
	L$_\ast$ (\Lsun) & 4 -- 30 & $\times 1.16$ & & Flat\\
	T$_\ast$ (K) & 4000 -- 5500 & 500 & & Flat\\
	M$_\rmn{d}$ (\Msun) & 10$^{-4.5}$ -- 10$^{-3}$ & $\times 10^{0.5}$& & Flat\\
	R$_\rmn{in}$ (AU)  & 0.3 -- 1.5 & 0.2 & & Flat\\
	R$_\rmn{out}$ (AU) & 200 & fixed & & \\
	h$_{100}$ (AU)  & 10 -- 35 & 5 & & Flat\\
	$\beta$  & 1.0 -- 1.3 & 0.05 & & Flat\\
	
	$i$ ($^\circ$) & 5 -- 85 & $\geq$2$^{\circ}$ & & $\sin i$\\
	A$_\rmn{V}$ (mag) & 0.1 -- 1.5 & 0.2 & & $HN (\sigma=0.3) $\\
	 \hline
	
	\end{tabular}
	\end{table}
		We seek to find a suitable fit by first building a database of $\sim 1.5\times10^6$\,single-star+disc models, the parameter space for which is shown in Table \ref{tab:gpars}. 
		These models cover a range based on previous literature investigations of the system where possible (e.g. L$_\ast$, T$_\ast$, M$_\mathrm{d}$) and a physically motivated range for the others.
		
		For each of these models, {\sc MC3D} returns output in the form of an SED from 0.3 to 3000\,$\mu \mathrm{m}$ and images covering the wavelength range of the MIDI observations.
		From each image, synthetic interferometric visibilities corresponding to the baselines used in the MIDI observations are calculated using fast Fourier transforms \citep{2007NewAR..51..565H}.
		The real and imaginary parts of the synthetic visibilities are determined by linear interpolation between adjacent values in the Fourier plane.

		To build models of the binary system, we combine pairs of models and compare the output to the MIDI visibilities and the SED of the system and its components.
		The fluxes used for the SED are given in Tab.~\ref{tab:sed}; where possible, we select the most recent and robust photometry available.  
		For each pair of models, we sum the fluxes at wavelengths where no observations are available which resolve the binary in order to robustly include the far-IR observations, while the resolved fluxes are compared to the respective member of the model pair.
		Due to the convergence of the resolved SEDs toward longer wavelengths through N\,\&\,Q bands, and the similarity of the SMA 
		fluxes, we consider a subset of binary configurations such that the two components differ only in inclination and position angles, and in which the NE component is obscured by its own disc.
		
		These simulated data are then used to calculate posterior-probability distributions and their marginal distributions for representative binary configurations under certain assumptions.
		A detailed description of these calculations is given in Appendix \ref{sec:bayes}.
		The likelihood for each datum is calculated using Eq.~\ref{eq:glike}, and no additional weighting is applied to the data beyond that implied by their uncertainties and bandwidth, treating photometry and interferometry identically.
		As a result, our fitting process is naturally most sensitive to the most precise data which integrate over the narrowest wavelength range, which is in general those between 7--13\,${\rm\mu m}$.
		The product of the likelihoods of all the data for a given model is then inserted into Eq.~\ref{eq:bayes} along with the product of the priors of the model (see Tab.~\ref{tab:gpars}) to give the posterior.
		The best-fitting model is the pair of single-star models which maximises the posterior.
		It is worth emphasising that any conclusions drawn in this way are valid only within the parameter space we consider. 
		
	\begin{table*}
	\caption{Data used for SED fitting}
	\label{tab:sed}
	\centering
	\begin{tabular}{l l l l l}
	\hline\hline
	Effective wavelength$^a$ ($\umu$m) & \multicolumn{3}{c}{Flux (Jy)}  & Ref. \\
	 & SW & NE & Unres & \\ \hline
	 & \\
	0.36 $\pm$ 0.025 & 0.0042 $\pm$ 0.00014 & $<$ 8$\times 10^{-5}$ $^b$ & ** & 1,2 \\
	0.43 $\pm$ 0.036 & 0.0081 $\pm$ 0.00067 & $<$ 0.00016 $^b$& ** & 1,2 \\
	0.55 $\pm$ 0.038 & 0.0217 $\pm$ 0.0008 & $<$ 0.0004 $^b$& ** & 1,2 \\
	0.7 $\pm$ 0.081 & 0.0510 $\pm$ 0.0007 & $<$ 0.001 $^b$& ** & 1,2 \\
	0.9 $\pm$ 0.097 & 0.117 $\pm$ 0.003 & $<$ 0.002 $^b$& ** & 1,2 \\
	1.25 $\pm$ 0.09 & 0.40 $\pm$ 0.06 & 0.014 $\pm$ 0.0098 & ** & 3 \\
	1.65 $\pm$ 0.12 & 0.94 $\pm$ 0.13 & 0.153 $\pm$ 0.0211 & ** & 3 \\
	2.16$\pm$ 0.13 & 1.63 $\pm$ 0.23 & 0.846 $\pm$ 0.117 & ** & 3 \\
	3.8 $\pm$ 0.22 & 1.88 $\pm$ 0.35 & 1.69 $\pm$ 0.311 & ** & 3 \\
	4.80 $\pm$ 0.2 & 3.50 $\pm$ 0.65 & 2.6 $\pm$ 0.47 & ** & 4 \\
	7.73 $\pm$ 0.245 & 13.7 $\pm$ 0.7 & 7.0 $\pm$ 0.4 & ** & 2 \\
	8.74 $\pm$ 0.269 & 14.4 $\pm$ 0.2 & 4.5 $\pm$ 0.1 & ** & 2 \\
	10.35 $\pm$ 0.357 & 15.0 $\pm$ 0.3 & 3.3 $\pm$ 0.1 & ** & 2 \\
	12.33 $\pm$ 0.415 & 20.5 $\pm$ 0.7 & 8.7 $\pm$ 0.3 & ** & 2 \\
	18.3 $\pm$ 0.527 & 28.4 $\pm$ 1.6 & 11.1 $\pm$ 0.8 & ** & 2 \\
	24.56 $\pm$ 0.699 & 35 $\pm$ 18 & 25.1 $\pm$ 13.5 & ** &  2\\
	70 $\pm$ 8.84  & ** &  ** & 55 $\pm$ 6 & 5\\
	100 $\pm$ 14.3 & ** &  ** & 80 $\pm$ 12 & 5\\
	160 $\pm$ 31.4 & ** &  ** & 66 $\pm$ 13 & 5\\
	450 $\pm$ 16.8 & ** &  ** & 12 $\pm$ 4 & 6\\
	850 $\pm$ 27.7 & ** &  ** & 2.0 $\pm$ 0.2 & 6\\
	870 $\pm$ 54.2 & ** &  ** & 1.66 $\pm$ 0.06 & 7\\
	1200 $\pm$ 164  & ** &  ** & 0.58 $\pm$ 0.06 & 8\\
	1270 $\pm$ 117 & ** &  ** & 0.469 $\pm$ 0.021 & 9\\
	1340 $\pm$ 22 & 0.1999 $\pm$ 0.0074 & 0.1701 $\pm$ 0.0083& ** & 10 \\
	 \hline
	 \multicolumn{5}{l}{$^a$ Defined as the central wavelength of the filter $\pm$ half of the smallest interval, symmetrical about }\\\multicolumn{5}{l}{the central wavelength, that contains at least 68.3\% of the area under the transmission curve.} \\
	 \multicolumn{5}{l}{$^b$ 3$\sigma$ upper limits} \\
	 \multicolumn{5}{l}{{\bf References: } (1) \citet{1981AJ.....86...62M}; (2) \citet{2011ApJ...729..145K} ; (3) Koresko (priv. comm.) ; } \\
	 \multicolumn{5}{l}{(4) \citet{2008poii.conf..519R} ; (5) \citet{2013A&A...551A..34S}; (6) \citet{2005MNRAS.357..975N};} \\
	 \multicolumn{5}{l}{ (7) \citet{2011ApJ...736..137S}; (8) \citet{1994A&A...291..546H}; (9) \citet{2003A&A...409..235C}; (10) this work}\\
	\hline
	\end{tabular}
	\end{table*}

\section{Discussion}\label{sec:dis}
	\subsection{Best-fitting model \& degeneracy}\label{sec:bfit}
\begin{figure*}
	\includegraphics[clip=true,trim=7cm 5cm 4cm 4cm,width=\textwidth,]{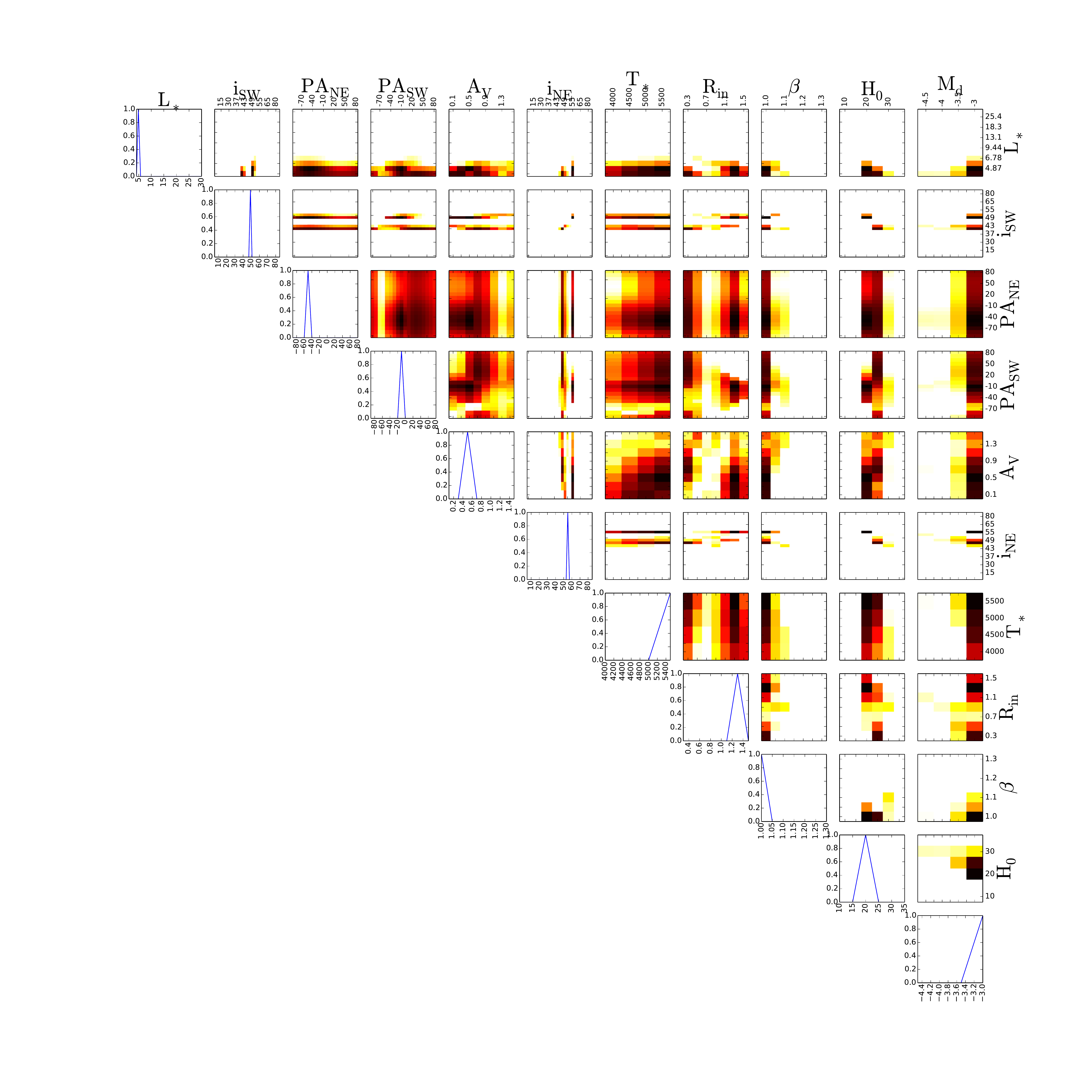}
	\caption{Cornerplot showing the degeneracies between the various free parameters in terms of the posterior-probabiltiy distribution marginalised over all other parameters, along with the marginal distributions of each free parameter. The colour scale is logarithmic, with each sub-plot separately normalised to its own maximum value; black indicates high probability and white low.}\label{fig:grid}
	\end{figure*}

	Table~\ref{tab:fit} lists the parameters of the best-fitting model, and Fig.~\ref{fig:grid} shows a corner plot of the marginalised posterior distributions, illustrating degeneracies between different input parameters.
	In particular, R$_{\rm in}$ is degenerate with the inclinations of both discs.
	These are bimodal, with models with smaller R$_{\rm in} \sim 0.3$ preferring solutions with inclinations $\sim 40 - 45^\circ$ rather than the $\sim 50 - 55^\circ$ solution of our best-fitting model.
	Otherwise, parameters are typically constrained to a small region near the value of the best-fitting model, or are upper or lower limits.
	Unfortunately, the grid resolution is not high enough to determine uncertainties from the marginalised distributions, however it is possible to derive certain robust constraints from Fig.~\ref{fig:grid}.
	Our results indicate that models with relatively low luminosity stars with massive, unflared, geometrically thick discs seen at intermediate inclinations are strongly preferred, while the stellar effective temperature and disc position angles are poorly constrained.
	The luminosity is lower than suggested by some previous models \citep[e.g.][]{1992ApJ...397..520W}.
	This is a result of the extreme near-IR variability of the NE component, whose K-band flux has varied by 2 orders of magnitude over the last 3 decades \citep{2011ApJ...729..145K}.

	\begin{table}
	\caption{Best-fitting model parameters}
	\label{tab:fit}
	\centering
	\begin{tabular}{l c}
	\hline\hline
	Parameter & Best Fit \\ 
	\hline
	 & \\
	L$_\ast$ (\Lsun) & 4.9 \\
	T$_\ast$ (K) & 5500\\
	M$_\rmn{d}$ (\Msun) & 10$^{-3}$\\
	R$_\rmn{in}$ (AU)  & 1.3  \\
	R$_\rmn{out}$ (AU) & 200 \\
	h$_{100}$ (AU)  & 20  \\
	$\beta$  & 1.00 \\
	
	$i_{\rm SW}$ ($^\circ$) & 49 \\
	$i_{\rm NE}$ ($^\circ$) & 55 \\
	A$_\rmn{V}$ (mag) &  0.5\\
    PA (SW) ($^\circ$)$^a$ &  -10 \\
	PA (NE) ($^\circ$)$^a$  &  -50 \\
	 \hline
	\multicolumn{2}{l}{$^a$ position angle east of north of the {\it minor} axis of the disc} \\
	\hline
	\end{tabular}
	\end{table}	
	
	\begin{figure*}
		\subfloat[]{\includegraphics[scale=0.475,clip=true,trim=0cm 0.cm 2.5cm 14.25cm]{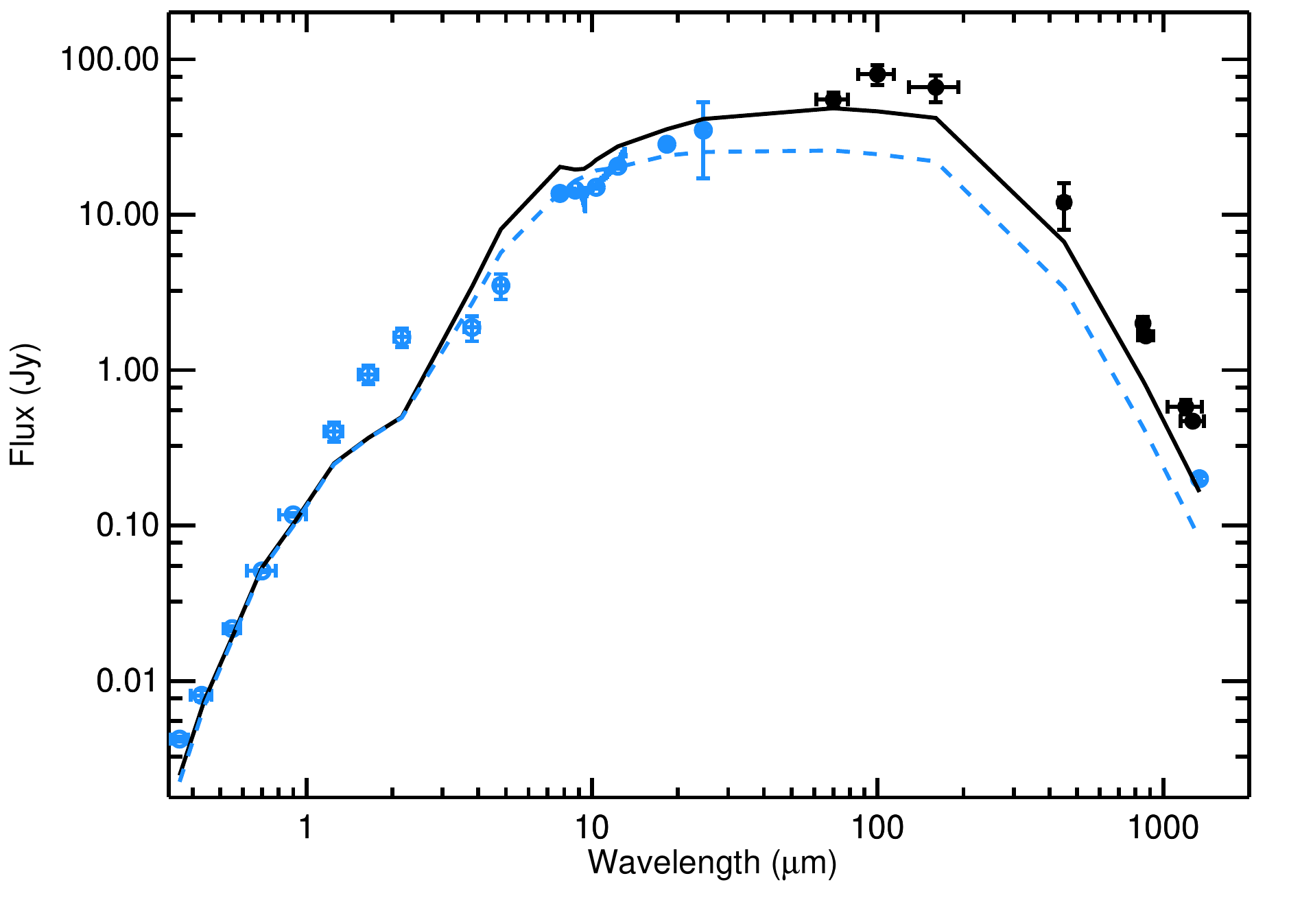}}	
	    \subfloat[]{\includegraphics[scale=0.475,clip=true,trim=0cm 0.cm 2.5cm 14.25cm]{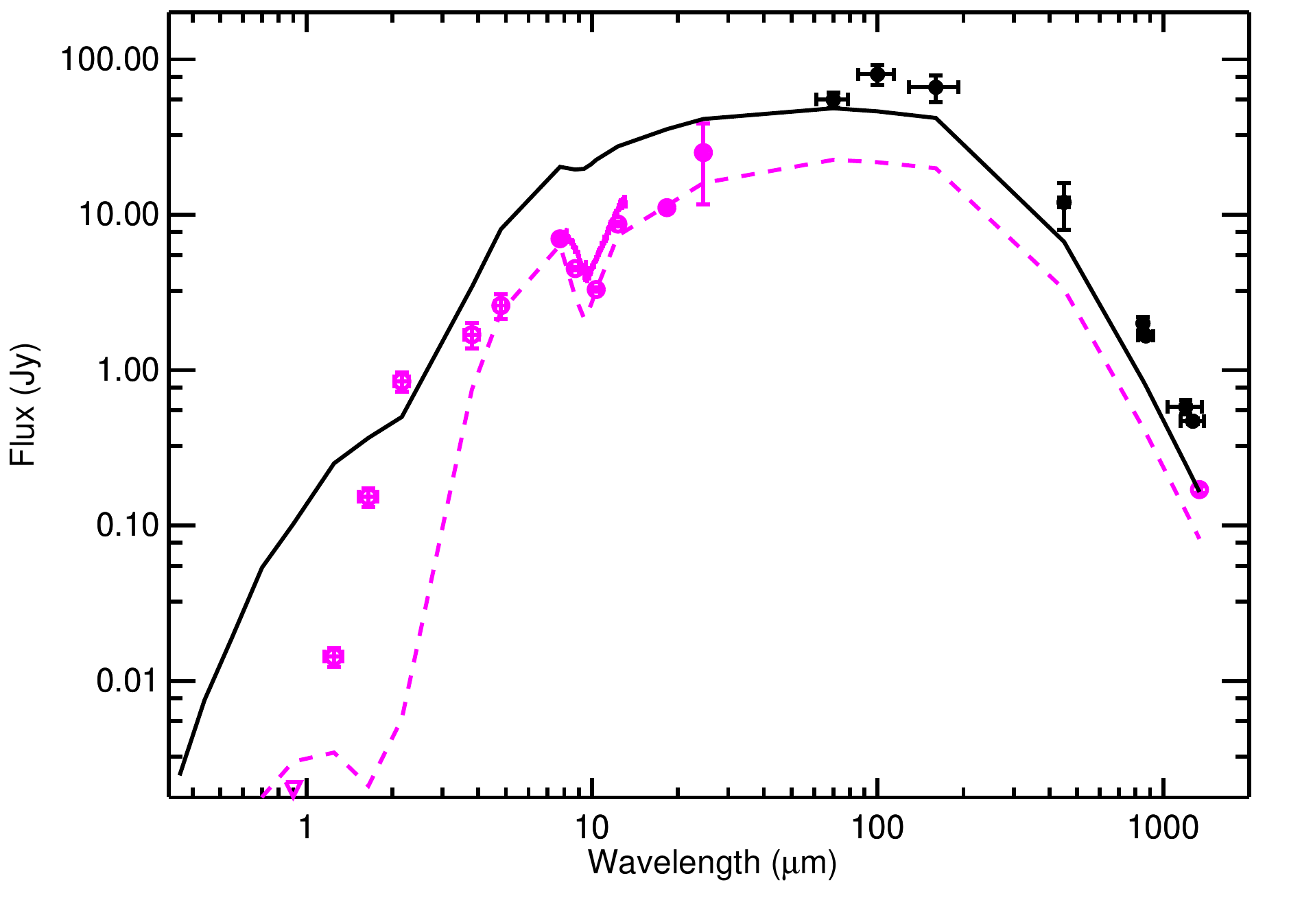}}	
	\caption{Simulated SEDs of the best-fitting model for the SW ({\it left}) and NE ({\it right}) components and the observed SEDs. In both cases, the solid black points correspond to photometry from observations which fail to resolve the two components of the binary (column {\it Unres} in Tab.\,\ref{tab:sed}), and the solid line corresponds to the sum of the model flux of both components. The coloured points correspond to resolved photometry, and the coloured dashed line indicates the best-fitting model's emission from each component separately. The triangle in the lower left of panel b indicates observed 3-sigma upper limit in I band; the upper limits at shorter wavelengths do not fit within the axes. }\label{fig:sed} 
	\end{figure*} 
	
	\begin{figure}
    \resizebox{\hsize}{!}{\includegraphics[scale=0.475,clip=true,trim=0cm 0.cm 2.5cm 14.25cm]{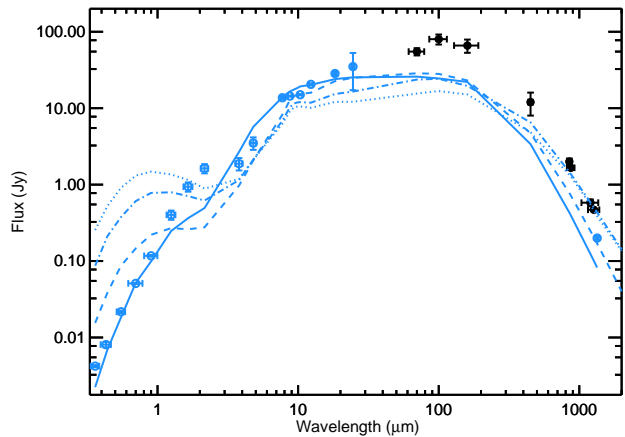}}	
	\caption{Influence of the maximum grain size on the SED of our best-fitting model for the SW component. As in Fig.~\ref{fig:sed}, the blue points correspond to the resolved observations of the primary, and the black points to the unresolved photometry. The solid line shows the SED for VV~CrA~SW shown in figure~\ref{fig:sed} with small grains only, the dashed, dot-dashed and dotted lines extend the size distribution to 2.5, 25 and 250~$\mu$m, respectively. The 2.5~$\mu$m grains fit the sub-mm flux and spectral index well, without significantly perturbing the mid -- far-infrared fluxes.} \label{fig:bigdust}
	\end{figure}
	
		\begin{figure}
    \resizebox{\hsize}{!}{\includegraphics[scale=0.5,clip=true,trim=1.cm 15.7cm 3.5cm 1cm]{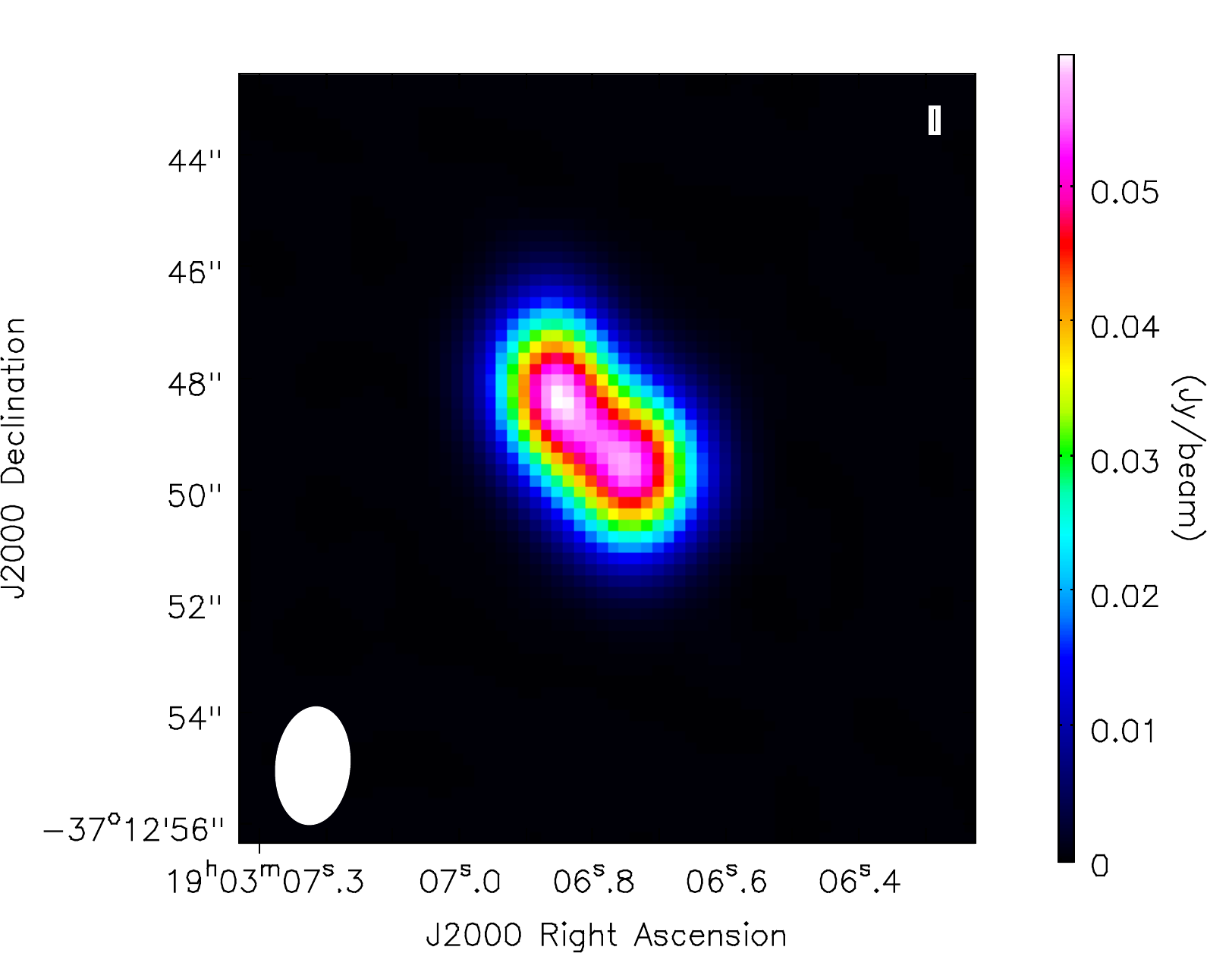}}	
	\caption{Simulated 1.3\,mm image based on our best-fitting model. Produced using CASA {\it simobserve}, neglecting all sources of noise other than UV-coverage. The total flux is underpredicted, but the discs are also too extended compared to Fig.\,\ref{fig:sma}} \label{fig:smacomp}
	\end{figure}
	
	\begin{figure*}
	\subfloat[]{\includegraphics[width=0.5\textwidth]{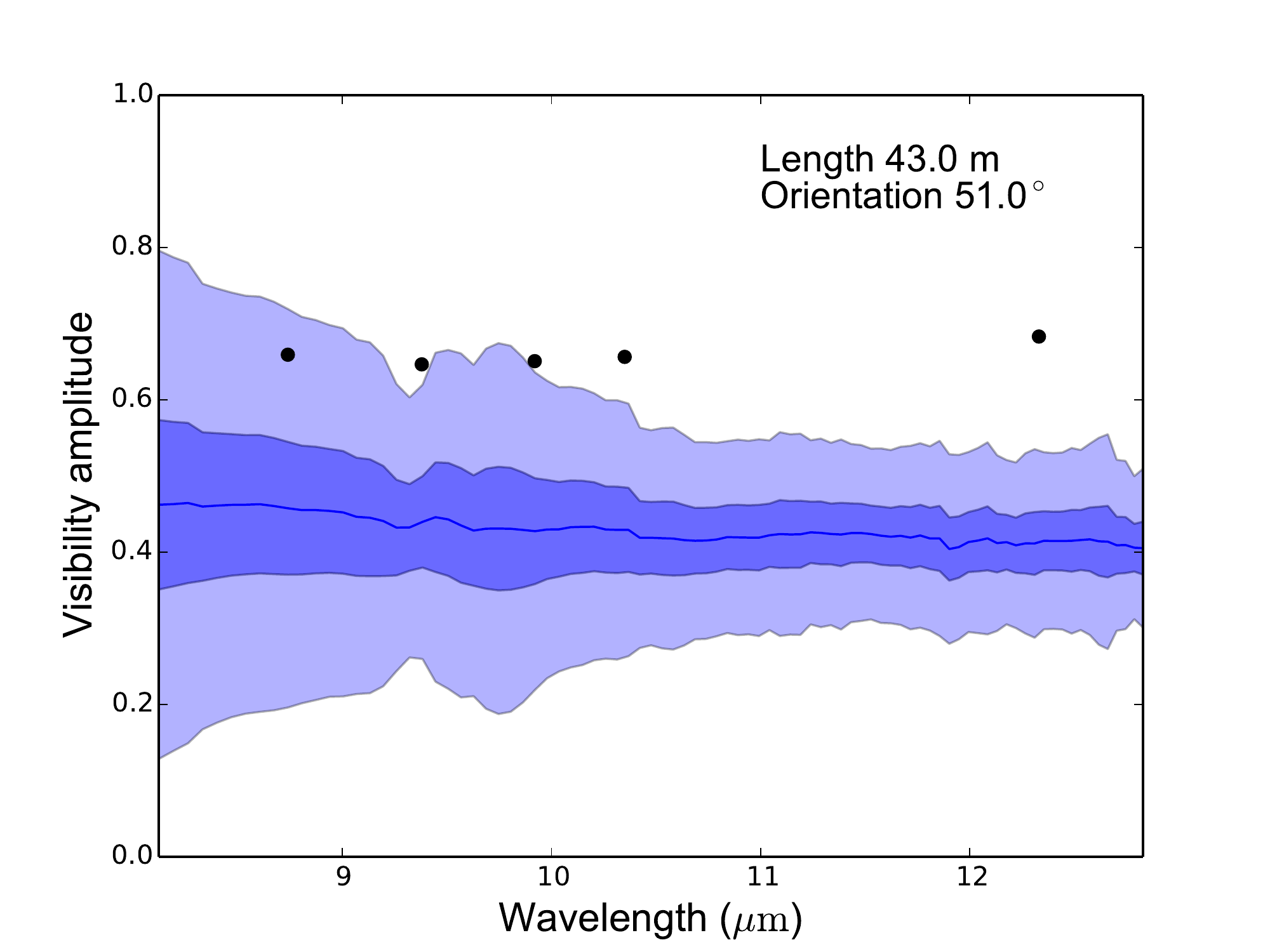}}
	\subfloat[]{\includegraphics[width=0.5\textwidth]{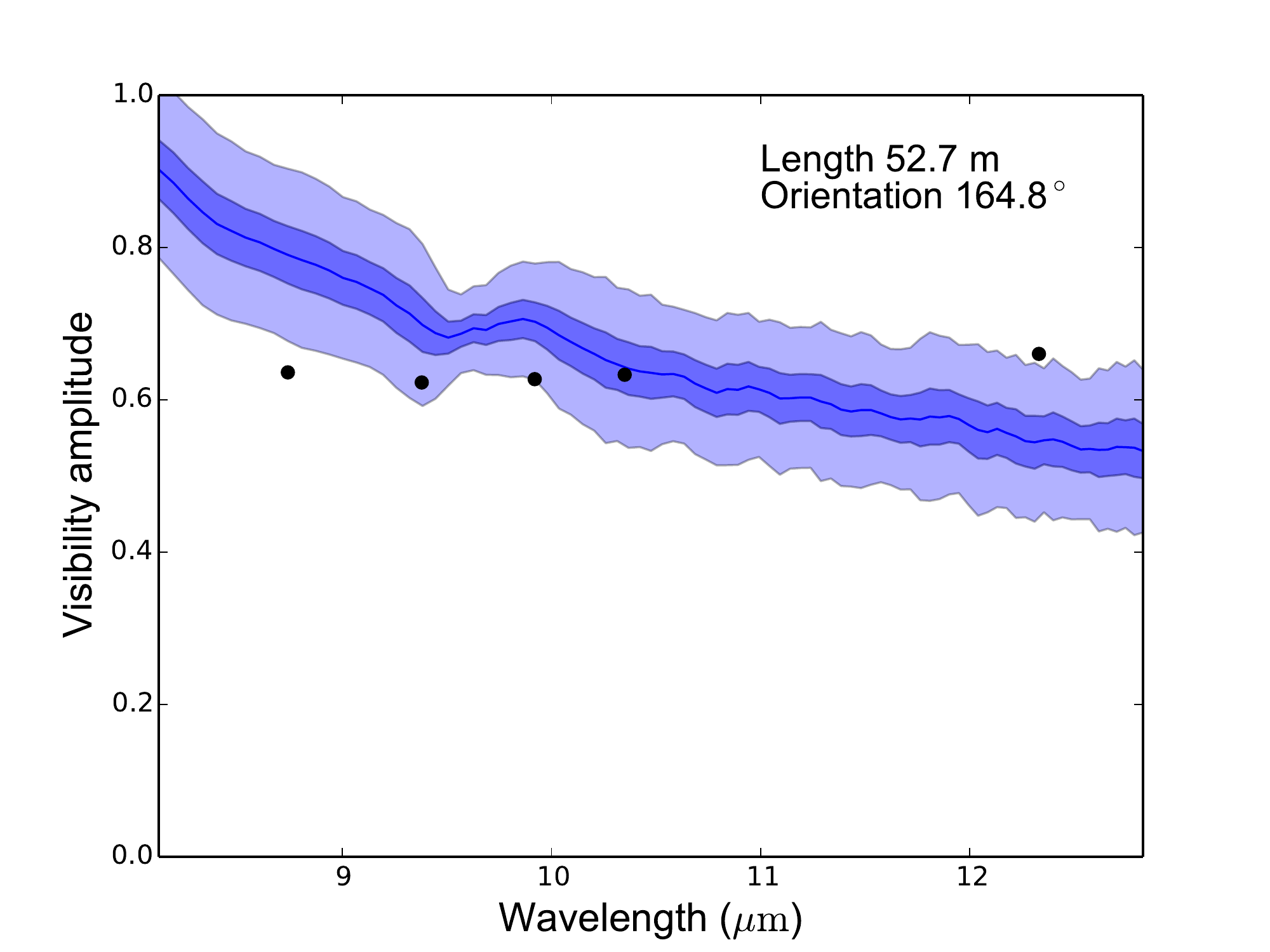}}
	
	\subfloat[]{\includegraphics[width=0.5\textwidth]{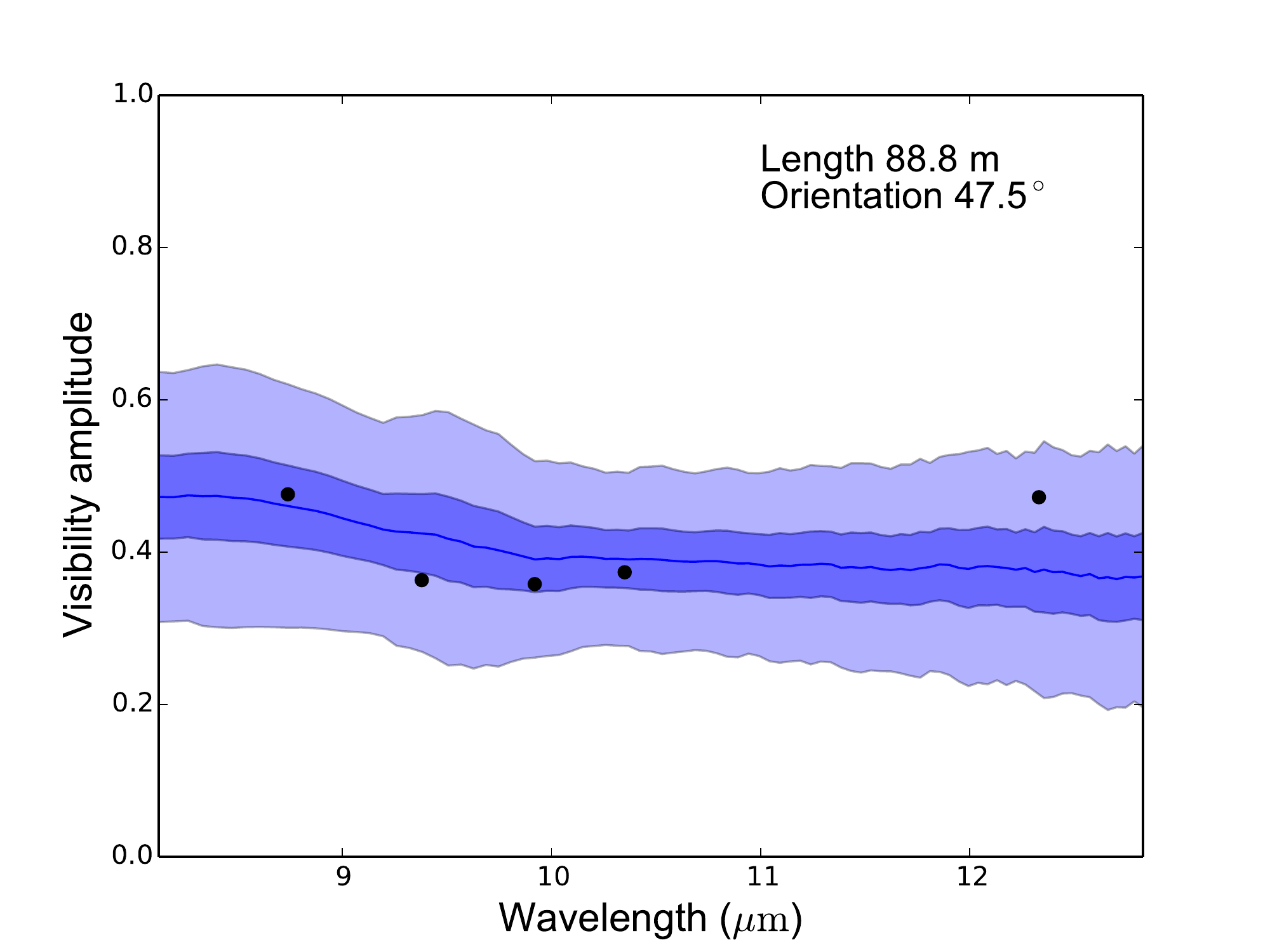}}
	\subfloat[]{\includegraphics[width=0.5\textwidth]{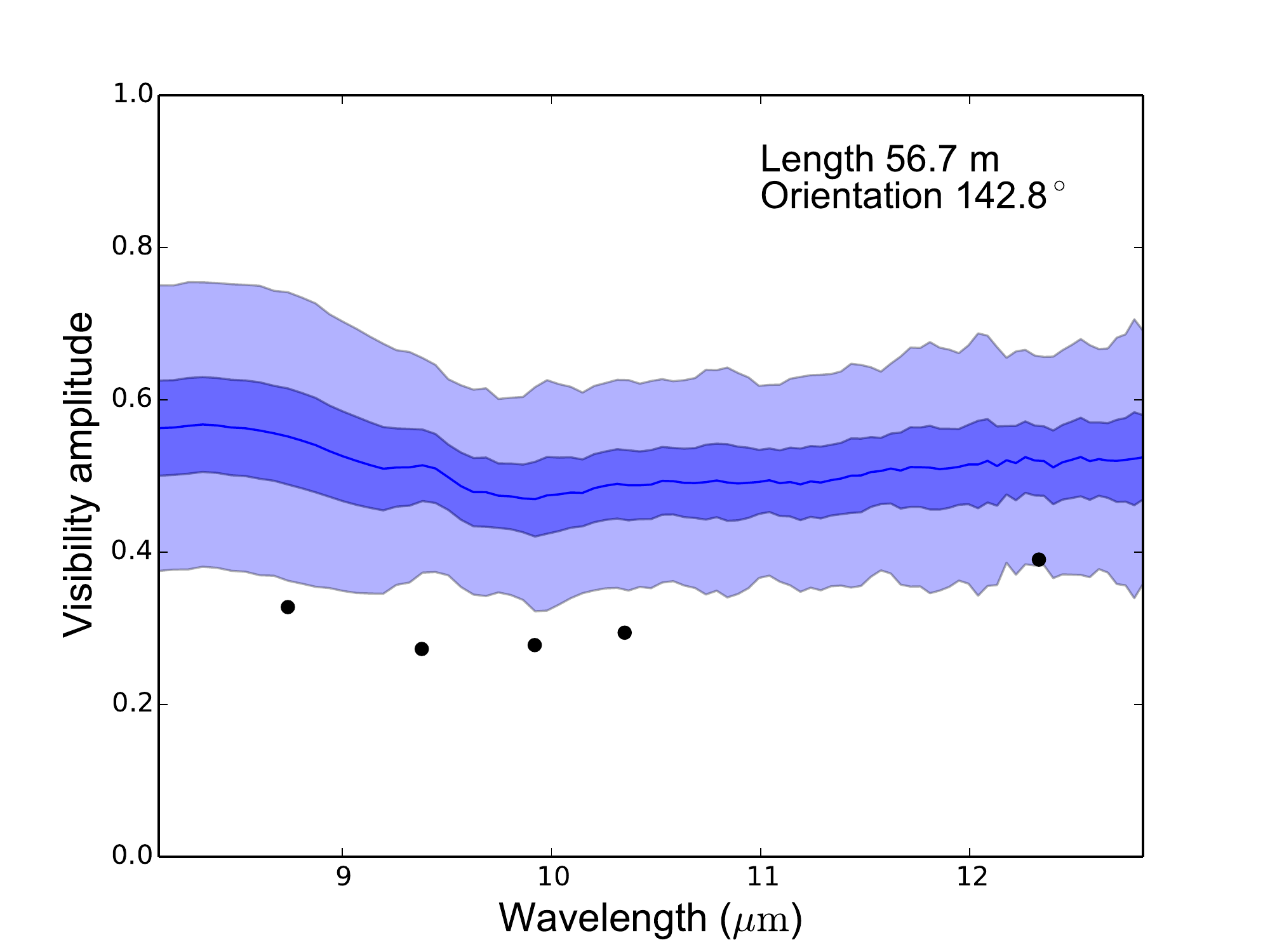}}
	\caption{Comparison of two selected observed visibilities and their counterparts in our best-fitting model. {\it a \& b:} Observations of the primary. {\it c \& d:} secondary component. The projected baseline length and orientation are given in the top-right corner of each plot. The full line indicates the observed visibility, the dark- and light-blue regions respectively indicate its 1- and 3-sigma regions, and the black points the simulated visibilities.}\label{fig:viscomp}
    \end{figure*} 		
	
	The SED of the best-fitting model is shown in Fig.~\ref{fig:sed}.
	Our best-fitting model reproduces the fluxes at short wavelengths ($\lambda\leq 30\mu$m) well, although the JHK-bands are underpredicted.
	However, despite the very high dust mass, it does not reproduce the FIR--sub-mm fluxes. 
	The grid does not include higher dust masses because, assuming a typical gas-to-dust ratio of $\sim 100$, the mass of the disc would then be comparable to that of the central object and self-gravity would begin to play a role \citep{2004MNRAS.351..630L,2005MNRAS.358.1489L}, and hence Eq.\,\ref{eq:rhodisc} would no longer adequately describe the density distribution.
	This cut-off coincides with an empirical upper limit to dust masses derived from sub-mm observations \citep{2013ApJ...773..168M}.
	Furthermore, the total dust mass of the binary implied by our model ($2\times 10^{-3}$\,\Msun) is 5\,times larger than that determined by \citet{2010A&A...515A..77L}. 
	As a result, further increases to the dust mass would not provide an acceptable solution.	
	
	Alternatively, increasing the maximum grain size would increase the sub-mm opacities, and hence increase the emergent flux at these wavelengths.
	This is demonstrated in Fig.~\ref{fig:bigdust}, which shows the effect on the SED of increasing the maximum grain size in steps of 1 dex from 0.25~$\mu$m to 250~$\mu$m.
	Increasing the maximum grain size leads to a significant increase in the emission at long wavelengths, and only begins to affect the quality of the fit in the MIR when a$_{\rm max}$ is very large.
	Even by increasing a$_{\rm max}$ by only one order of magnitude, the fit to the sub-mm data is dramatically improved, giving good agreement with the observed fluxes and spectral index.
	However, these models assume that the grain-size distribution is the same throughout the disc, with no consideration of grain settling or radial drift.
	As described in \S~\ref{sec:dust}, we lack the observational constraints required to justify the complexity that this would add to our model, and must therefore accept that a larger dust mass is required to reproduced the observed (sub-)mm emission.
	However, the compact nature of the 1.3\,mm emission provides a further hint that grain-growth and settling processes should be included in future models.
	Figure~\ref{fig:smacomp} shows a simulation of the SMA observations computed using our best-fitting model as input to the {\it simobserve} \& {\it simanalyze} tasks in CASA.
	As noted, the model does not reproduce the flux of the observations, being $\sim$ 2--3 times lower.
	However, the core of the emission of each disc appears more extended in the simulated image than the observation, suggesting a component of compact emission is missing from our model.
	A likely explanation is the lack of large dust grains in our model; these grains should rapidly settle to the midplane and experience radial drift, making them more centrally concentrated.
	As large grains are more efficient mm-wave emitters than the ISM grains in our model, this would result in an additional component of compact emission, which may dominate the SMA observations.
	
	Figure~\ref{fig:viscomp} and Appendix~\ref{sec:viscomp} compare the MIDI observations with the visibilities of our best-fitting model.
	In general, the model fits well, but tends to predict worsening resolution at longer wavelengths, in contrast to the observed visibilities, which tend to decrease with increasing wavelength.
	The emitting area of the disc as a function of wavelength must therefore increase faster compared to the size of the beam than our model assumes.
	One plausible explanation is an effect of the scattered flux; if $\sim 1\mu$m dust grains are present at the disc surface as implied by infrared spectroscopy of sources in nearby star-forming regions \citep{2006ApJS..165..568F,2009ApJ...703.1964F,2006ApJ...639..275K,2010ApJS..188...75M,2010ApJ...714..778O}, these would efficiently scatter the radiation emitted by the disc, producing substantial additional extension.
	However, RT modelling of the optically thin case using {\sc MC3D} suggests that the reduction in temperature of larger grains outweighs this effect, reducing the effective size of the disc. 
	However, dust temperatures in real discs can be substantially different from the optically thin case, but this case is not feasible in the current version of {\sc MC3D}.
	There is no evidence of emission from PAHs or stochastically heated grains, which would also introduce extended MIR emission.
	While is is possible that our model has converged on a solution with too little flaring, none of the observed visibilities show the rise around 12\,$\mu$m seen by \citet{2005A&A...441..563V} for flared discs, arguing against this.
	Indeed, the observations are broadly consistent with their models for self-shadowed discs, suggesting that, rather than flaring, there may be additional structure missing from our model.
	The binary nature of the system makes structure a logical conclusion, as the discs would be subjected to perturbations by the companions;
	given the qualitative similarity to RW~Aur \citep[see e.g.][and references therein]{2015MNRAS.449.1996D}, VV~CrA may be an interacting binary at an earlier stage of evolution. However, we do not expect any signatures of interaction to be visible in existing data -- any tidal tail would likely be lost in the missing flux in the SMA image, and the MIDI data lack the {\it u--v} coverage required to detect spiral waves in the inner discs. 
	These possibilities could plausibly be tested using the new generation of XAO imagers (e.g. SPHERE) to acquire near-infrared imaging polarimetry, or by imaging the thermal emission at very high resolution for example with ALMA or MATISSE, discussed in further detail below.

	Based on this tension between data and models, it is clear that future studies must consider more complex models of VV~CrA.
	As noted above, there are suggestions of grain growth, decoupling between gas and dust, and additional structure in the discs.
	At present it is not feasible to explore the full parameter space of these effects and their possible interactions with detailed radiative transfer models, and as noted above, existing data are unable to constrain such modelling, necessitating additional observational constraints. 
	It is likely that models including such additional physics would produce a somewhat different best-fitting model than the simple parametrisation used here.
	However, we expect this to have a larger effect on the disc mass, flaring index, and inclination angles, while the inner radius, low luminosity and foreground extinction should be relatively robust.
	
	\subsection{Mass \& evolutionary state}
	
	We can determine an approximate stellar mass $M$ and age $t$ by comparing the temperature and luminosity of the best-fitting model to evolutionary tracks.
	Using the tracks of \citet{1997MmSAI..68..807D}, we construct interpolating functions for $T_\ast\left(M,t\right)$ and $L_\ast\left(M,t\right)$ from the isochrones, which are then solved simultaneously for $M$ and $t$. 
	We thus determine a mass of 1.8\,\Msun\,and an age of 3.0\,Myr; if the accretion luminosity determined in \S\ref{sec:spec} is subtracted from the luminosity of the best-fitting model, we arrive at 1.7\,\Msun\,and an age of 3.5\,Myr. 
	Similar estimates are found using other evolutionary models, for example \citet{1999ApJ...525..772P} and \citet{2000A&A...358..593S}.
	These values are inconsistent with previous estimates which inferred a very young T Tauri system based on the shape of the SED in the mid-infrared \citep[e.g.][]{1997ApJ...474..455P}, but approximately consistent with the model of \citet{1997ApJ...480..741K} for the primary. 
	However, radiative-transfer-model fitting is only weakly sensitive to the stellar temperature, and relatively small variations imply significant changes in the mass and age.
	Only by improving spectral-type determinations can matters be improved.
	
	Our best-fitting model implies a stellar radius of 2.4\,\Rsun for the primary, which allows us to estimate an accretion rate from the accretion luminosity given in \S\ref{sec:spec} as in \citet{1998ApJ...492..323G} and \citet{2008A&A...478..779S}.
	Assuming a mass of 1.7\,\Msun\,as determined above and a truncation radius of 5\,R$_\ast$, we derive an accretion rate of 4.0 $\times 10^{-8}$\,\Msun\,yr$^{-1}$.
	This is similar to the values of $\dot{M} / M$ seen in spectroscopic surveys of nearby star-forming regions \citep[e.g.][]{2014A&A...561A...2A,2015arXiv151008255M}.

	\subsection{Outlook}
	As it is clear that a variety of questions remain concerning the state of VV~CrA, we will now discuss which observations should be targeted to allow future modelling efforts to improve upon our limitations.
	As the major shortcoming of our models is the failure to consider larger dust grains, resolved observations in multiple ALMA bands should be considered a priority.
	Spatially resolved spectral-index maps of the two discs would allow a reasonable prescription of grain growth and settling to be incorporated in future models, giving more reasonable estimates of the dust masses.
	A by-product of such observations would be high-resolution spectral line data, allowing the disc inclinations and stellar masses of both components to be directly constrained.
	Knowing the masses of the stars would allow for a substantially improved age determination, while directly measured inclinations could conclusively differentiate between the two cases in \citet{2009ApJ...701..163S}.
	
	Another aspect that would considerably enhance modelling efforts would be robust knowledge of the spectral types of the stars and their accretion rates.
	To date, this has proven nearly impossible due to the near-complete veiling of the photospheric lines in the primary, and the lack of flux from the secondary.
	The most promising course of action seems to be deep, high-resolution near-infrared spectroscopy covering a large range of wavelengths, 
	or perhaps combined optical-NIR spectroscopy from instruments like XSHOOTER; observations of this kind would maximise the chances of resolving weak photospheric lines.
	Combining such data with a fitting method as in \citet{2013A&A...558A.114M} would allow the incorporation of the effect of veiling, improving the chances of finding a suitable solution.
	
	As noted above, it is likely that our model does not include sufficient structure to correctly describe the disc.
	Probing such structures on scales $\lesssim$\,10\,AU requires milli-arcsecond resolution, which is now possible thanks to extreme adaptive optics and interferometry.
	The new generation of planet-hunting imagers (e.g. SPHERE, GPI, SCExAO)  provide precisely the high-resolution capabilities required in scattered light, while ALMA can achieve the same resolution in the sub-mm, where optical depths are much lower; these may also be able to detect larger-scale structures, such as a circumbinary disc, envelope or tidal tail.
	Meanwhile, MATISSE will provide sufficient uv-coverage in the mid-infrared to facilitate interferometric imaging in a reasonable amount of observing time, and will achieve a similar angular resolution to both ALMA and SPHERE, albeit with a much smaller field-of-view. 
	Previous studies \citep{2014A&A...572L...2R} have shown that structures are rarely detectable at all wavelengths simultaneously, so the ability to cover a large wavelength range is critical.	

\section{Conclusions}\label{sec:conc}
	We have conducted a detailed study of the enigmatic pre-main-sequence binary system VV~CrA.
	New near-infrared spectroscopy of the primary has derived an accretion luminosity of 0.81~$\pm$~0.07~\Lsun, and placed an upper limit on the extension of Br\,$\gamma$ emission in the N--S direction of 10\,mas.
	Archival SMA observations have been used to produce the first resolved image of the binary at 1.3\,mm, and mid-infrared interferometry probes the inner regions of both discs.
	
	Based on these observations and data from the literature, we have presented a model for VV~CrA, showing that even a small misalignment of the discs is a plausible explanation for the differences between the two components of the binary, with inclinations of 50--55$^\circ$.
	The MIDI observations allow us to break the degeneracies inherent in SED fitting, improving the constraints on the mutual inclination.
	This model suggests a lower luminosity for the central stars than previous work, most likely a result of the variability of the infrared companion.
	Based on our model, we infer an age of 3.5\,Myr and stellar masses of 1.7\,M$_\odot$; combining these results with the accretion luminosity yields an accretion rate of 4.0 $\times 10^{-8}$\,\Msun\,yr$^{-1}$.
	
	We also suggest observational priorities that may resolve outstanding questions. 
	While resolved ALMA observations should be the priority to resolve disc structures and allow future models in include grain growth and settling, there is also a strong case for further near-infrared observations of both components.
	In particular, determination of at least two of mass, spectral type and luminosity for both components would provide a robust test of evolutionary models.
	Similar studies of other pre-main-sequence binaries should be conducted in order to build up a sample of coeval systems to constrain models of star \& planet formation and evolution.

\section{Acknowledgments}

We would like to thank the anonymous referee for their comments, Casey Deen for performing the CRIRES observations and reading the manuscript, Chris Koresko for providing Keck/NIRC photometry, Hongchi Wang for helpful analysis of NACO photometry, and Peter Abraham for helpful discussions.
This reseach was funded under DFG programme nos. WO 857/10-1 \& WO 857/13-1.
CFM gratefully acknowledges an ESA Research Fellowship.
The Submillimeter Array is a joint project between the Smithsonian Astrophysical Observatory and the Academia Sinica Institute of Astronomy and Astrophysics and is funded by the Smithsonian Institution and the Academia Sinica.
\bibliographystyle{mn2e_trunc8} 
\bibliography{vvcra-midi}

\appendix
\section{Bayesian inference}\label{sec:bayes}

This discussion is based heavily on \citet{2010arXiv1008.4686H} and \citet{2010arXiv1009.2755A}, which the interested reader should peruse for a deeper understanding of the methods involved.

Suppose one has a set of data $D$ which contains a number of observations, and wishes to identify which model $M_i$, drawn from a set of $n$ models, fits these observations best.
In a frequentist approach, one would seek to quantify the quality of the fit either by minimising the $\chi^2$ statistic or by maximising the (logarithmic-)likelihood function $\mathcal{L} = P\left(D \vert M_i\right)$ i.e. by considering only how well the model reproduces the dataset.

However, this does not account for any previous knowledge about the object being modelled, whether from statistical, theoretical or literary considerations.
Failing to include this information in the fitting process may lead one to prefer a model with slightly higher likelihood in spite of its intrinsically lower probability.
To give an example appropriate to the main content of this article, consider a number of randomly oriented discs, where inclination $i=0$ implies a face-on disc.
For geometrical reasons, the probability of $\pi/3 \leq i \leq \pi / 2$ is one half, and hence higher inclinations are intrinsically preferred.
Therefore, if a low-inclination model has higher $\mathcal{L}$ in a particular case, the increased likelihood must outweigh the intrinsically lower probability of this configuration for the low-inclination model to be considered the better fit.

Such \textit{a priori} information can be encoded as the distribution of the probabilities of the models $P\left( M_i \right)$ which is referred to as the \textit{prior probability distribution} or more often simply the prior.
Given that what one wishes to calculate is the probability of model i given the observations or $P\left(M_i \vert D\right)$ (the \textit{posterior}) one quickly arrives at

\begin{equation}
P\left(M_i \vert D\right) = \frac{P\left(D \vert M_i\right) P\left( M_i \right)}{P\left(D\right)},
\label{eq:bayes}
\end{equation}

which it can be clearly seen is Bayes theorum.
Correct evaluation of the posterior therefore depends upon the correct choice of priors and likelihood functions, and their correct evaluation.

In the case of astronomical photometry, the statistical uncertainties are dominated by Poisson noise and are therefore nearly gaussian, and the choice of likelihood function is straightforward. 
If observation $d_j$ belongs to our datased D, and consists of a flux $F_j$ observed at wavelength $\lambda_j$ with associated photometric uncertainty $\sigma_{F,j}$ and bandwidth $\sigma_{\lambda,j}$ and the covariances are zero\footnote{In all likelihood, this is not a very good assumption, since the effective wavelength of a filter during an observation depends upon the spectral shape of the emission being observed.}, then provided that the modelled wavelengths agree with those observed the likelihood can be calculated from:
\begin{equation}
\mathcal{L}_{ij} = \frac{\sigma_{F,j} \sigma_{\lambda,j}}{2\pi} \exp\left(\frac{-\sigma^2_{\lambda,j}}{2\sigma^2_{\lambda,j}\sigma^2_{F,j}}\left( F_j - F_{M_i,j}\right)^2 \right)
\label{eq:glike}
\end{equation}

where $F_{M_i,j}$ and $\lambda_{M_i,j}$ are the wavelength and flux produced by model $M_i$ that correspond with observation $d_j$.

Although the distribution of uncertainties for other astronomical observables are in general \textit{not} gaussian \citep[e.g.][]{1994A&A...287..676R}, in the absence of further information, the assumption of gaussianity is no worse than any other.
Hence, it is common to apply Eq. \ref{eq:glike} to all observables.
For example, in the case of interferometric observations, the fluxes are replaced with the visibilities, and the bandwidth with the spectral resolution.

In the case of observables where two parameters must be fitted, equation \ref{eq:glike} must be altered slightly.
For example, in the case of a polarimetric observation, where both the polarisation fraction $p$ and angle $\theta$ must be fitted, it becomes

\begin{align}
	\mathcal{L}_{ij} = \frac{\sigma_{p,j} \sigma_{\theta,j}}{2\pi} \exp\left(\frac{-1}{2\sigma^2_{\lambda,j}\sigma^2_{F,j}} \right.	
	&\quad\left( \sigma^2_{p,j}\left(p_j - p_{M_i,j}\right)^2 \right. \\
	&\quad\left.\left. +  \sigma^2_{\theta,j}\left(\theta_j - \theta_{M_i,j}\right)^2 \right) \frac{}{}\right). \nonumber
\end{align}

To find the likelihood of model $i$, one must then calculate the product $\mathcal{L}_{i} = \prod\limits_j \mathcal{L}_{ij}$

Once the posterior has been appropriately evaluated, the best fit can be extracted from the maximum of the distribution.
However, one is usually more interested in the \textit{credible interval} of each variable, usually taken as the 68.3\% credible region (or {\it 1--sigma region}).
This is the region within which 68.3\% of parameter estimates lie.
To determine this, one must calculate the marginal distributions of the posteriors by integrating over all other variables i.e. for two variables $a$ and $b$:
\begin{equation}
P(M_i \vert D, a) = \int\nolimits_{b^\prime} P(M_i \vert D, a, b)\, \mathrm{d}b^\prime,
\end{equation}

which leaves one with a one-dimensional distribution of the posterior.
From this, the 1--sigma region can be determined by integrating the distribution to find the region that contains 68.3\% of the probability.
This region is non-unique, so it is common to define it either as the narrowest region, as a region symmetrical about the median, or one symmetrical about the mean.
The results above refer to the narrowest region.

\section{Simulated and observed visibilities}\label{sec:viscomp}
FIgures \ref{fig:fullviscompSW} \& \ref{fig:fullviscompNE} show the full set of observed MIDI visibilities and the simulated visibilities for our best-fitting model.

	\begin{figure*}
	\subfloat[]{\includegraphics[width=0.5\textwidth]{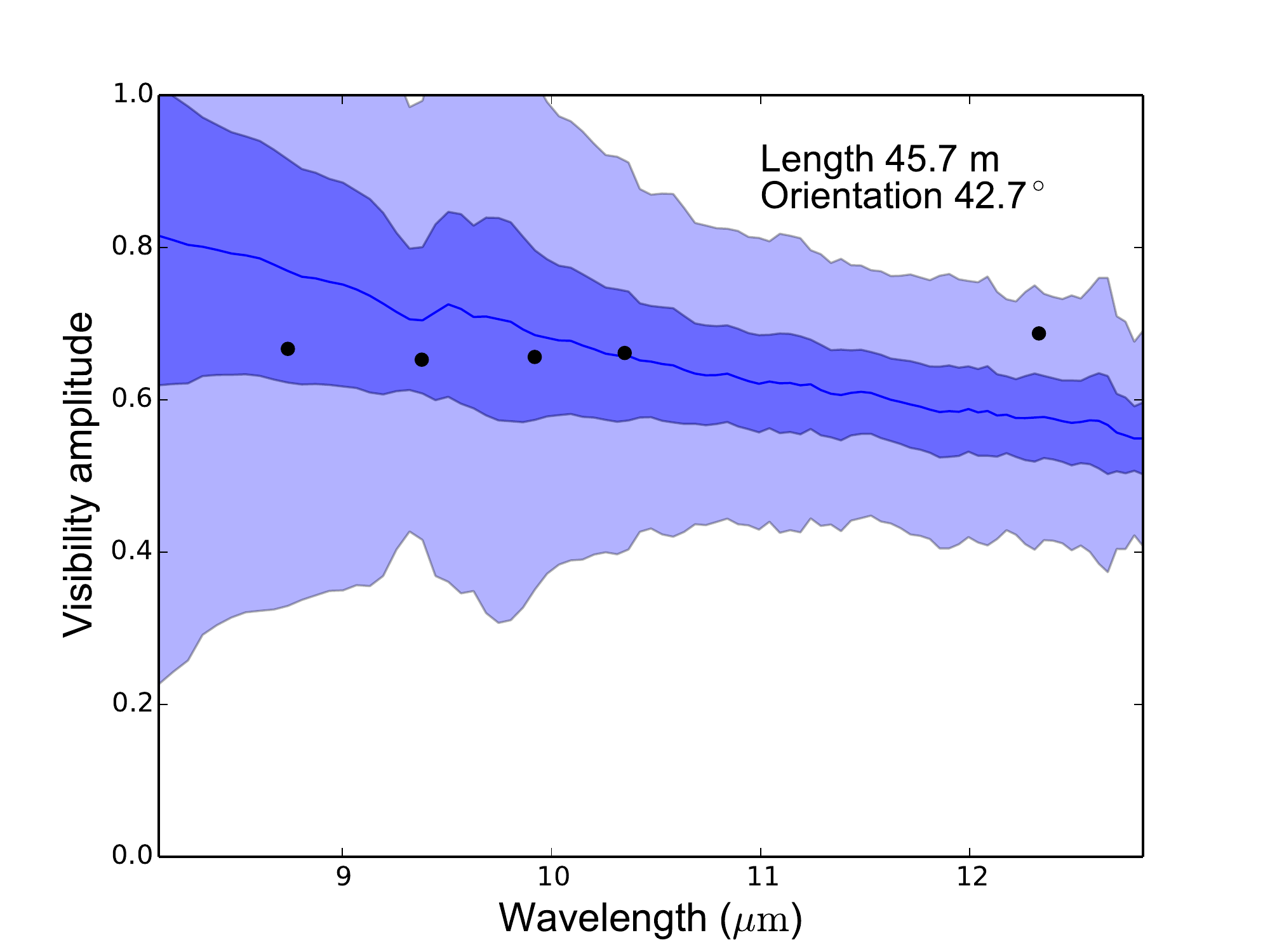}}
	\subfloat[]{\includegraphics[width=0.5\textwidth]{SWvis_1.pdf}}\\
	\subfloat[]{\includegraphics[width=0.5\textwidth]{SWvis_2.pdf}}
	\subfloat[]{\includegraphics[width=0.5\textwidth]{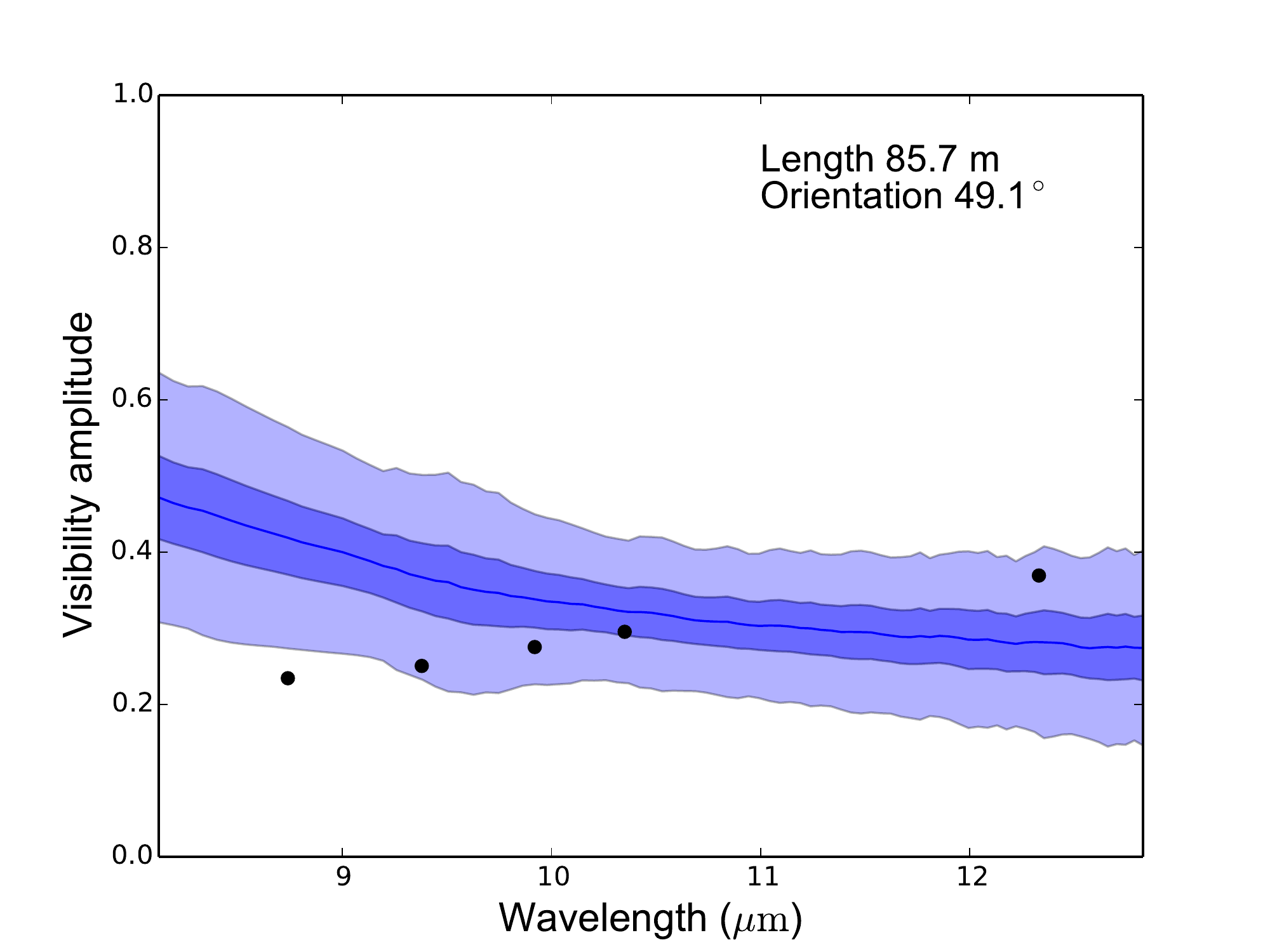}}\\
	\subfloat[]{\includegraphics[width=0.5\textwidth]{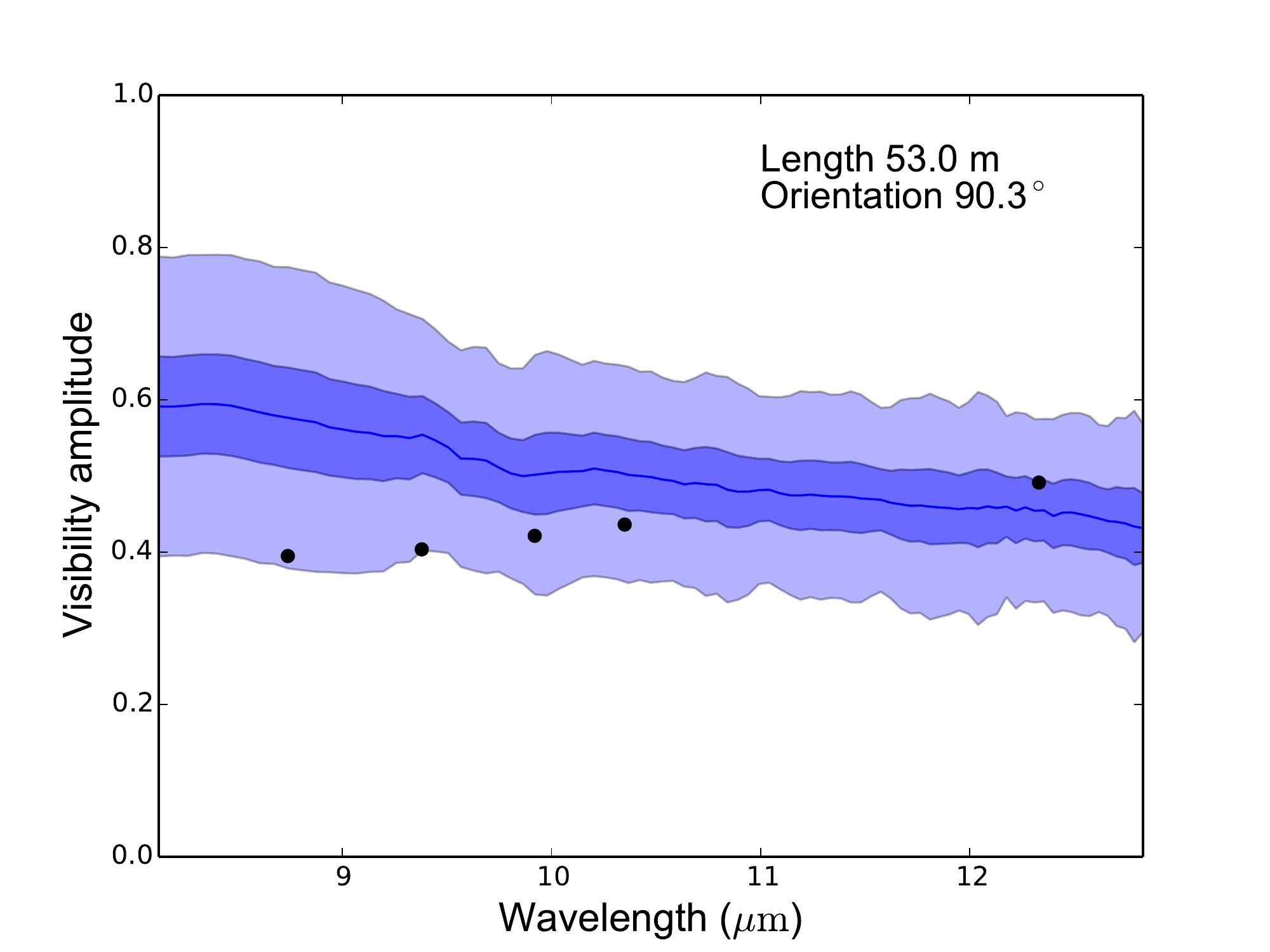}}
	\subfloat[]{\includegraphics[width=0.5\textwidth]{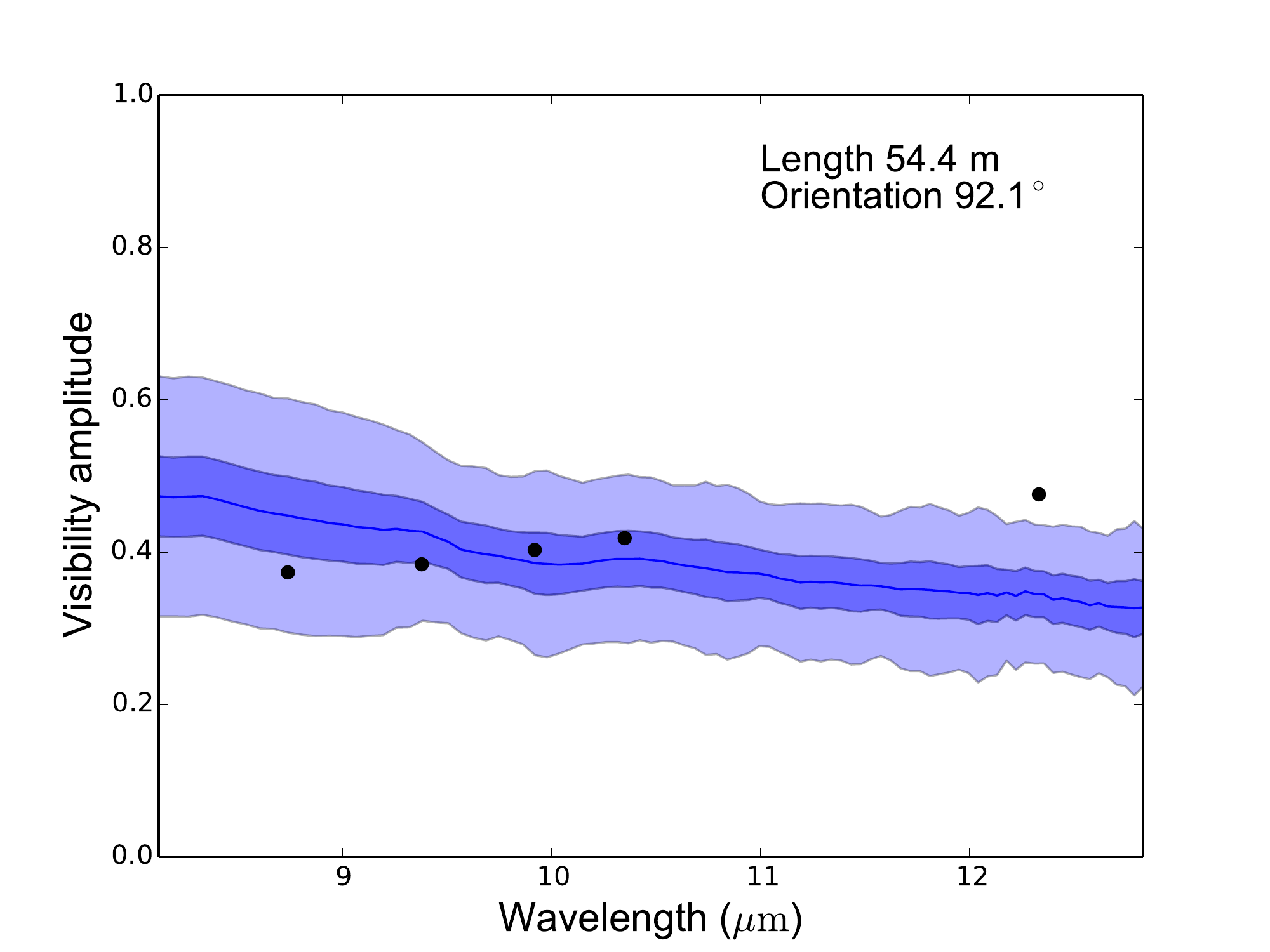}}\\
	\caption{Comparison of observed visibilities and their counterparts in our best-fitting model as in Fig.~\ref{fig:viscomp} showing all observations for the primary.} 
	\addtocounter{figure}{-1}
	\label{fig:fullviscompSW}
	\end{figure*}
	\begin{figure*}
	\subfloat[]{\includegraphics[width=0.5\textwidth]{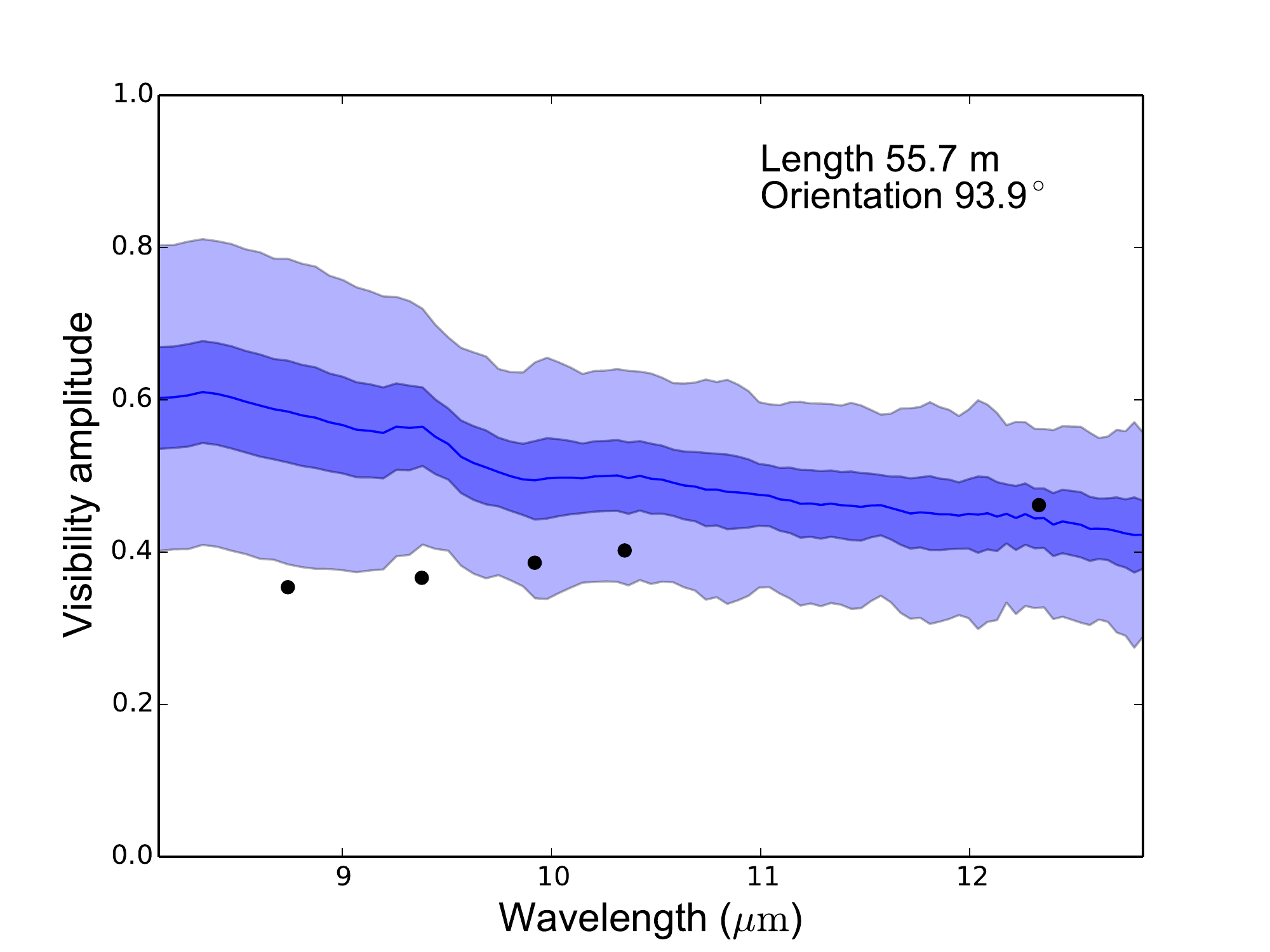}}
	\subfloat[]{\includegraphics[width=0.5\textwidth]{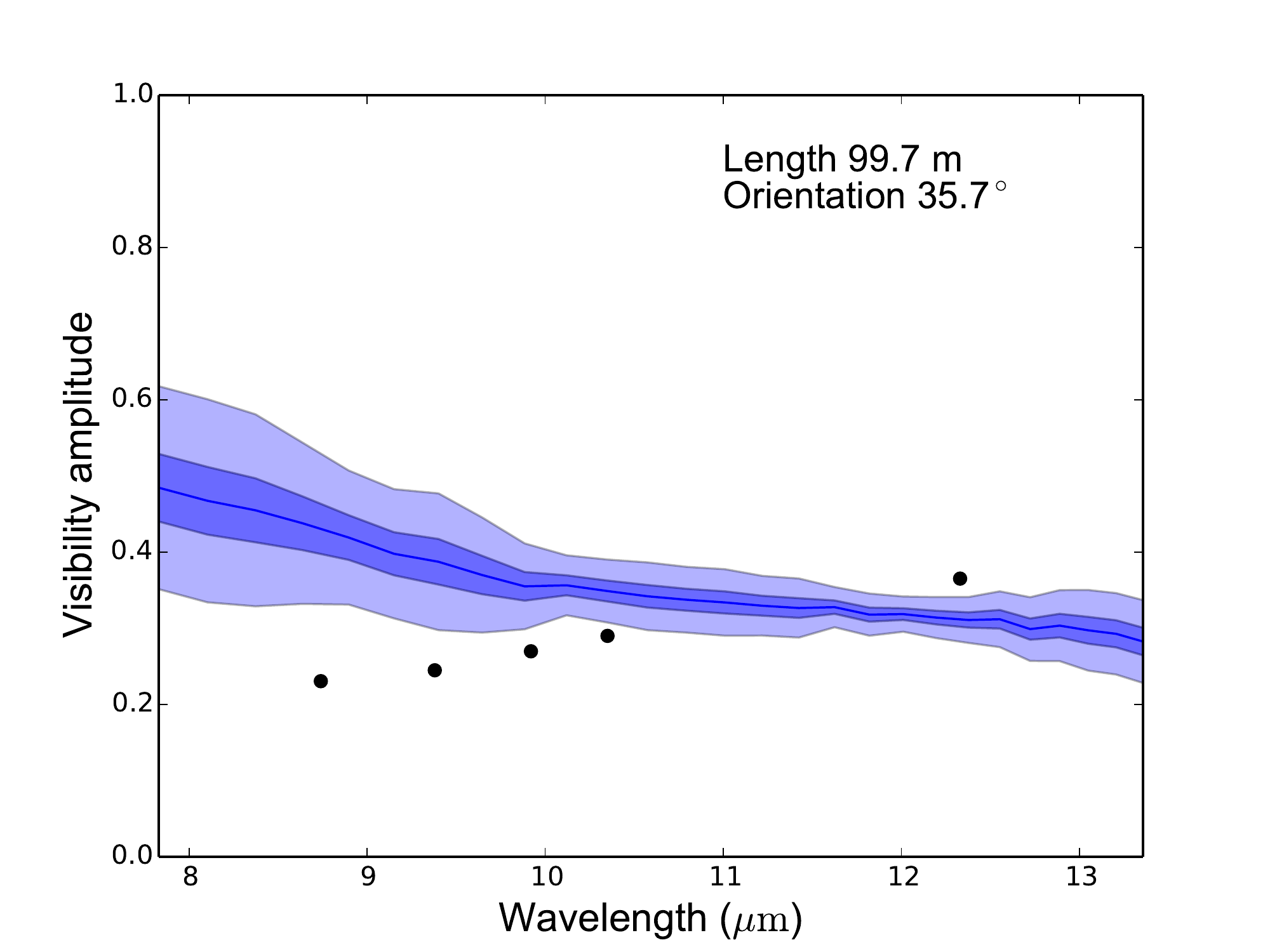}}\\
	\subfloat[]{\includegraphics[width=0.5\textwidth]{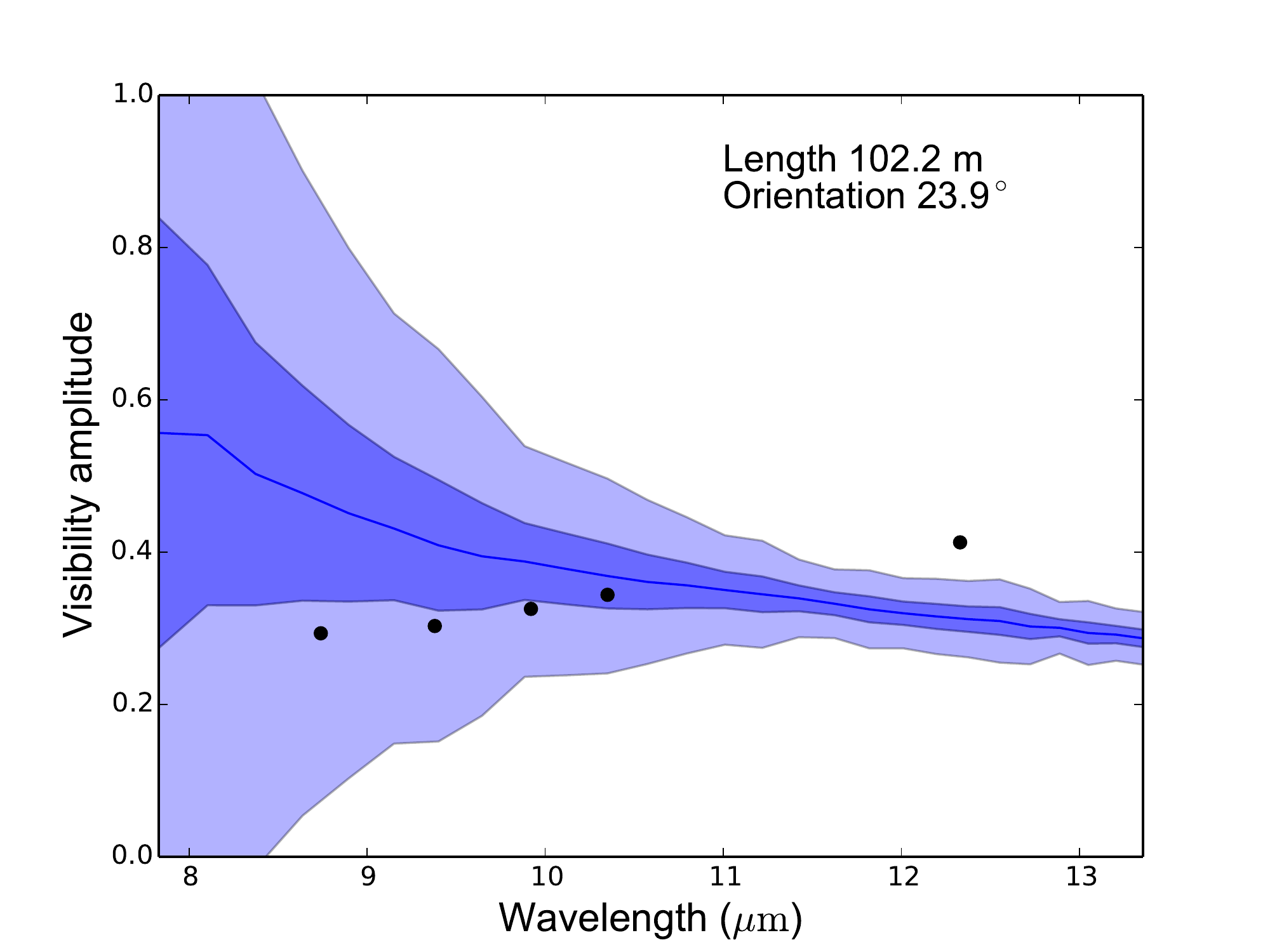}}
	\caption{{\it cont.}}
	\end{figure*}
	\begin{figure*}	
	\subfloat[]{\includegraphics[width=0.5\textwidth]{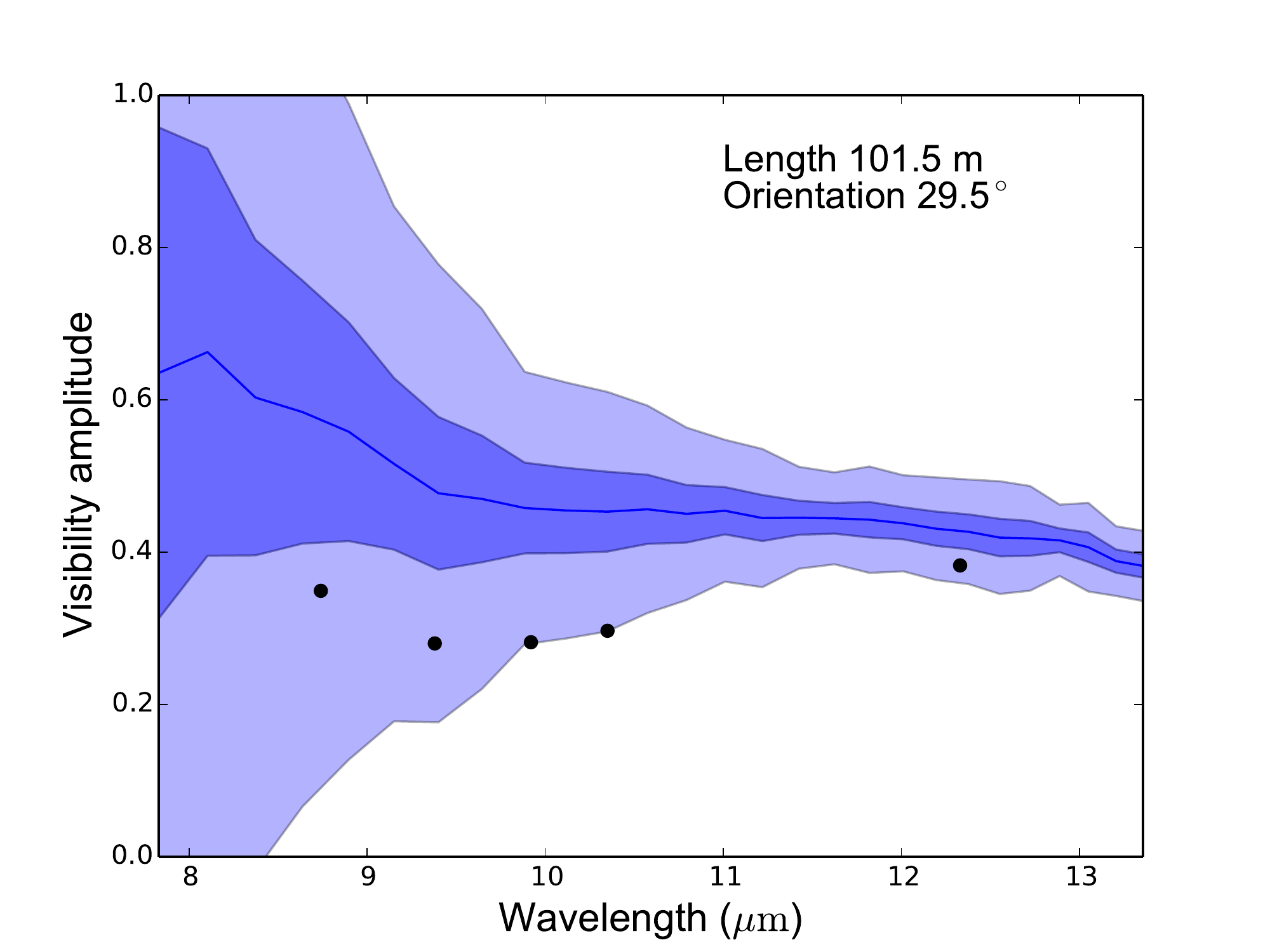}}
	\subfloat[]{\includegraphics[width=0.5\textwidth]{NEvis_1.pdf}}\\
	\subfloat[]{\includegraphics[width=0.5\textwidth]{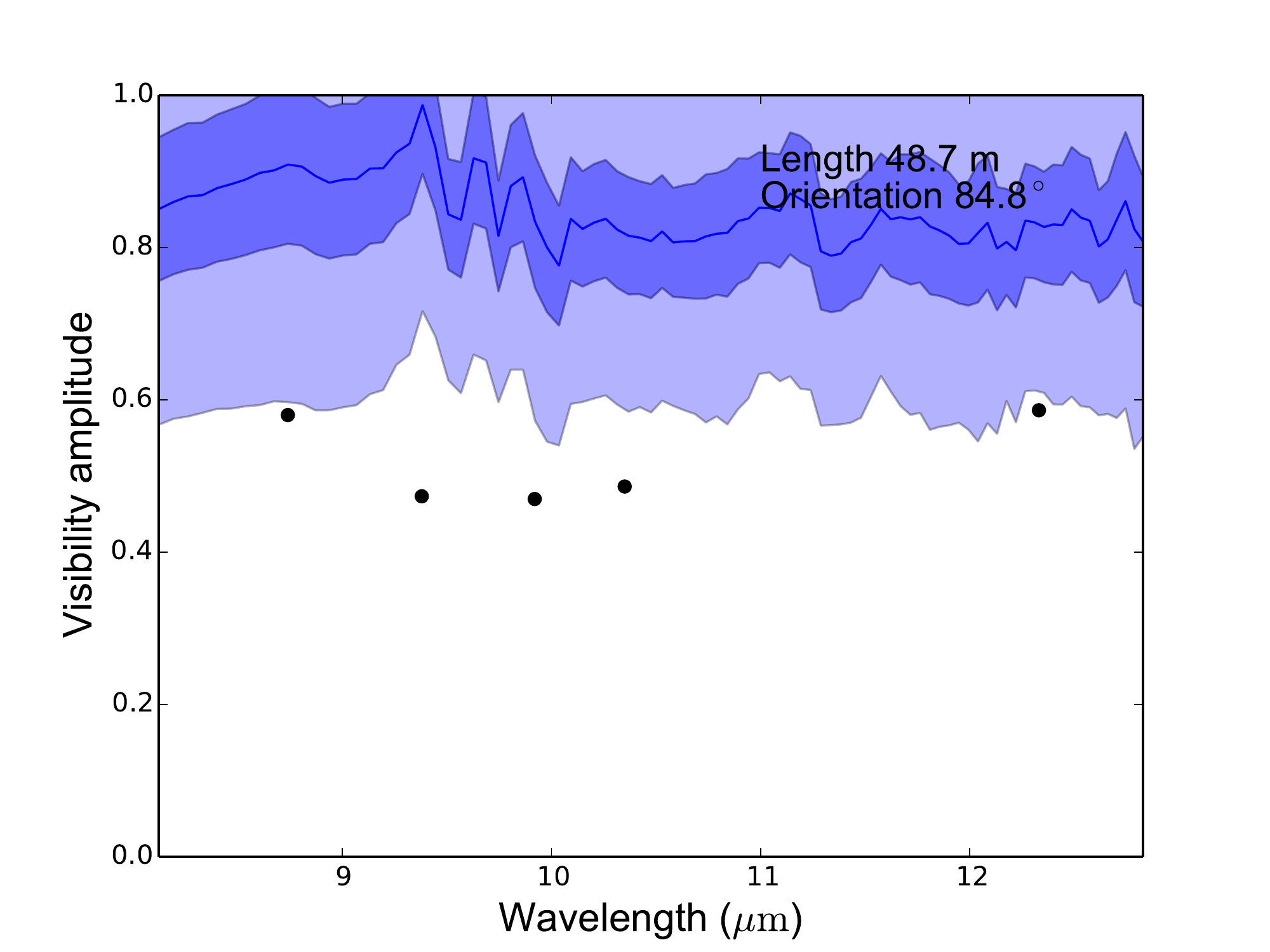}}
	\subfloat[]{\includegraphics[width=0.5\textwidth]{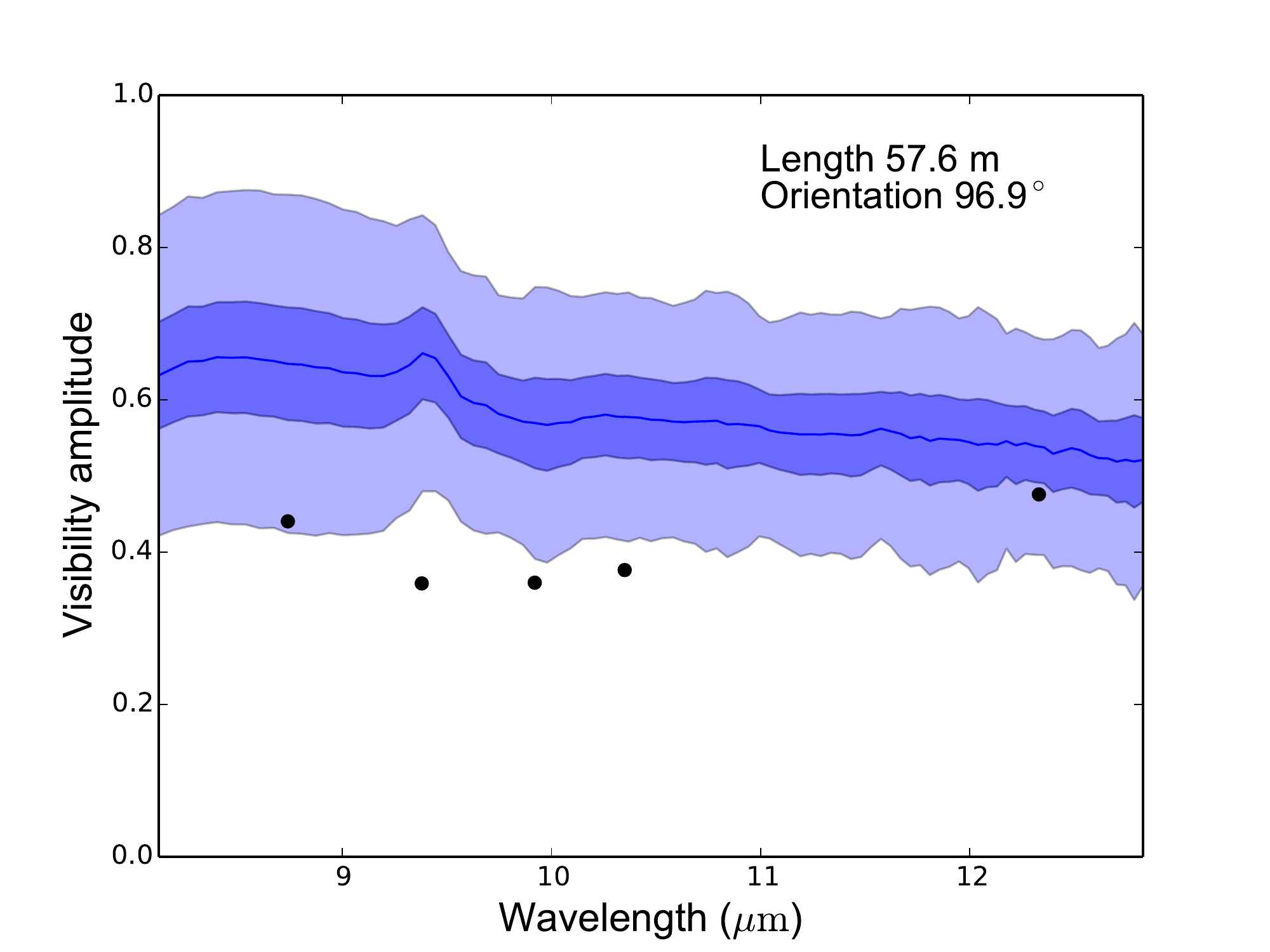}}\\
	\subfloat[]{\includegraphics[width=0.5\textwidth]{NEvis_4.pdf}}
	\caption{As Fig. \ref{fig:fullviscompSW}, showing data for the secondary.} 
	\label{fig:fullviscompNE}
    \end{figure*}

\end{document}